\documentclass[lettersize,journal]{IEEEtran}
\usepackage{amsmath,amsfonts}
\usepackage{algorithm}
\usepackage{array}
\usepackage[caption=false,font=normalsize,labelfont=sf,textfont=sf]{subfig}
\usepackage{textcomp}
\usepackage{stfloats}
\usepackage{url}
\usepackage{verbatim}
\usepackage{graphicx}
\usepackage{cite}
\usepackage{nomencl}
\makenomenclature%
\usepackage{etoolbox}
\usepackage{verbatim}
\usepackage{array}
\usepackage{multirow}
\usepackage{longtable}
\usepackage{rotating}
\usepackage{threeparttable}
\usepackage{eurosym}
\usepackage{makecell}
\usepackage{soul, color, xcolor}

\let\labelindent\relax 
\usepackage{enumitem}

\usepackage{algpseudocodex}
\usepackage{float}
\usepackage{amsmath,amssymb,amsfonts}

\begin{document}

\title{Multi-timescale Trading Strategy for Renewable Power to Ammonia Virtual Power Plant in the Electricity, Hydrogen, and Ammonia Markets}

\author{Sirui Wu,~\IEEEmembership{Student Member,~IEEE,} Jin Lin,~\IEEEmembership{Member,~IEEE,} Jiarong Li, Feng Liu,~\IEEEmembership{Senior Member,~IEEE,} Yonghua Song,~\IEEEmembership{Fellow,~IEEE,} Yanhui Xu,~\IEEEmembership{Member,~IEEE,} Xiang Cheng,~\IEEEmembership{Student Member,~IEEE,} and Zhipeng Yu,~\IEEEmembership{Student Member,~IEEE}

\thanks{This study is supported by the State Key Laboratory of Alternate Electrical Power System with Renewable Energy Sources (Grant No.LAPS22021).}
\thanks{S. Wu, J. Lin, J. Li, F. Liu, X. Cheng, and Z. Yu are with the Department of Electrical Engineering, Tsinghua University, Beijing, 100087, China.}
\thanks{Y. Song is with the Department of Electrical and Computer Engineering, University of Macau, Macau 999078, China, and also with the Department of Electrical Engineering, Tsinghua University, Beijing 100087, China}
\thanks{Y. Xu is with State Key Laboratory of Alternate Electrical Power System with Renewable Energy Sources (North China Electric Power University), Beijing 102206, China.} 
}

\markboth{Journal of \LaTeX\ Class Files,~Vol.~14, No.~8, August~2021}%
{Shell \MakeLowercase{\textit{et al.}}: A Sample Article Using IEEEtran.cls for IEEE Journals}


\maketitle

\begin{abstract}
    Renewable power to ammonia (RePtA) is a prominent zero-carbon pathway for decarbonization. Due to the imbalance between renewables and production energy demand, the RePtA system relies on the electricity exchange with the power grid. Participating in the electricity market as a virtual power plant (VPP) may help to reduce energy costs. However, the power profile of local photovoltaics and wind turbines is similar to those in the market, resulting in rising energy costs under the conventional strategy. Hence, we develop a multi-timescale trading strategy for the RePtA VPP in the electricity, hydrogen, and ammonia markets. By utilizing the hydrogen and ammonia buffer systems, the RePtA VPP can optimally coordinate production planning. Moreover, we find it possible to describe the trading of electricity, ammonia, and hydrogen in a unified framework. The two-stage robust optimization model of the electricity market is extended to multiple markets and solved by the column and constraint generation (CC\&G) algorithm. The case is derived from an actual project in the Inner Mongolia Autonomous Region. Sensitivity analysis demonstrates the economic advantages of an RePtA VPP joining multiple markets over conventional strategy and reveals the necessity of the hydrogen and ammonia buffer and reactor's flexibility.

\end{abstract}

\begin{IEEEkeywords}
Renewable power to ammonia (RePtA), Virtual power plant (VPP), Multiple timescales, Multimarket trading strategy.
\end{IEEEkeywords}

\section{Introduction}\label{Introduction}

    \IEEEPARstart{R}{enewable} power to ammonia (RePtA) is a prominent zero-carbon pathway for the decarbonization of the chemical industry\cite{c01}\cite{c02}. It has gained increasing interest globally\cite{c03}\cite{c04}. The first commercial-scale green ammonia plant is planned in Western Jutland, Denmark, and designed to produce more than 5000 t/y of green ammonia from renewable power\cite{c05}. \cite{c06} shows that flexible green ammonia production based on renewable energy can enable a competitive levelized cost of ammonia (LCOA) to $\$$483/t in Taltal, Chile. The 'Asian Renewable Energy Hub' project in Pilbara, Australia, is expected to produce 9 million tonnes of ammonia per year based on 26 GW of renewable energy\cite{c07}. In addition, countries such as the UAE, Saudi Arabia, Oman, and Qatar are also actively supporting renewable energy to ammonia projects\cite{c08}.
    
    However, due to the multi-timescale imbalance between local renewables and production energy demand, the RePtA system relies on electricity exchange with the power grid \cite{c09}. On a monthly timescale, renewable generation varies with the seasons, while the energy demand follows the agricultural cycle because 85 \% of ammonia is utilized for fertilizer\cite{c10}. On a daily timescale, renewables have inherent intermittency. 
        However, due to the current constraints of the chemical process \cite{c43}\cite{c44}, the power to produce ammonia and hydrogen remains stable in the chemical factories in China.
    Such differences between the upstream supply and downstream demand result in power imbalances on multiple timescales. Previous studies of the RePtA system have mainly used a fixed tariff for the electricity from power grid\cite{c11}\cite{c12}, but this has been demonstrated to be uneconomical\cite{c13}.

    Participating in the electricity market as a virtual power plant (VPP) may benefit the RePtA system to obtain affordable electricity. However, energy costs might rise when conventional operating schedules are adopted. 
       In a specific region, the PV plants and wind turbines tend to be built in areas where renewable resources are most abundant. For example, in the Inner Mongolia Autonomous Region of China, most renewable energy generators are located in Baotou. In other words, the same type of generators are affected by the weather in a similar way.
    Therefore, a local power shortfall could always occur when there is an energy shortage in the market. Similarly, when there is an energy surplus in the RePtA VPP, there is also excess energy in the market, which is unfavorable for elecctricity sales. This is a challenging engineering issue that negatively affects the competitiveness of the RePtA system.

    By utilizing the hydrogen and ammonia buffer systems, the RePtA VPP can optimally coordinate production planning. The hydrogen and ammonia buffer systems include the buffer tanks and reactors.
    
        Since hydrogen and ammonia buffer tanks can store up to several weeks' worth of stock\cite{c14}\cite{c15}, load-shifting on multiple timescales is possible.
    
    Furthermore, increasing reactor's flexibility also helps to reduce energy costs. A shorter power adjustment interval helps the RePtA VPP to respond more swiftly to price fluctuations.

    Existing works on a the VPP’s trading strategy mainly focus on a single electricity market and only consider the spot market\cite{c16}. These studies examine trading strategies to maximize daily profit\cite{c17}\cite{c18} and the stochastic nature of renewable energy is usually considered\cite{c19}\cite{c20}. The optimization problems are usually represented by a MILP model\cite{c21} and solved using algorithms such as the branch-and-bound algorithm\cite{c22}, enhanced particle swarm optimization algorithm\cite{c23}, and genetic algorithms\cite{c24}. \cite{c25} determines the optimal control schedules of the controllable devices of the VPP based on the direct load control algorithm. \cite{c26} integrates EVs and wind production to participate in the day-ahead (DA) electricity market. Distribution network constraints are incorporated into the VPP bid optimization model in \cite{c27}. \cite{c28} proposes a real-time smart energy management model for a VPP using a multiobjective, multilevel optimization-based approach.

    However, such strategies are not suitable for the proposed RePtA VPP. First, unlike the flexible electrical loads commonly found in these studies (e.g., air conditioners, electric vehicles), the energy demand in the RePtA VPP comes from multimarket trading. The hydrogen and ammonia transactions entail corresponding operational constraints at different timescales. Moreover, the RePtA VPP always joins the long-term tradings. In addition to the spot market, transactions in futures markets need to be considered.

    Hence, we develop a multi-timescale trading strategy for the RePtA VPP in the electricity, hydrogen, and ammonia markets. The contributions of this paper are as follows:
    
    \begin{enumerate}[label = \arabic*)] 
    
    \item Under the conventional strategy without load-shifting, the RePtA VPP suffers from uneconomical trading prices. Therefore, we propose a multi-timescale RePtA VPP trading strategy in the electricity, hydrogen, and ammonia markets. By utilizing the hydrogen and ammonia buffer systems, the RePtA VPP can optimally coordinate production planning across hours and weeks to reduce energy costs.

    \item By modeling transactions across the three markets, we are able to describe the trading of electricity, ammonia, and hydrogen in a unified framework. Based on this finding, the two-stage robust optimization model of the electricity market is extended to multiple markets, which can be solved by the column and constraint generation (CC\&G) algorithm. Receding horizon optimization is utilized to articulate the multiple timescales.

    \item The case is derived from a project in the Inner Mongolia Autonomous Region, China. According to the sensitivity analysis, a hydrogen buffer is more effective for hourly price fluctuations, while an ammonia buffer is more practical for longer-termscale price fluctuations. Furthermore, reducing the adjustment period of the ammonia synthesis reactor (ASR) from 14 days to 1 day results in an 7.0\% reduction in LCOA.

    \end{enumerate}

    The remainder of the paper is organized as follows: the motivation for the RePtA VPP trading strategy is introduced in Section \uppercase\expandafter{\romannumeral2}. Section \uppercase\expandafter{\romannumeral3} creates the operation and trading model. In Section \uppercase\expandafter{\romannumeral4}, a multimarket trading strategy for the RePtA VPP is proposed considering stochastic renewable energy. Case studies are presented in Section \uppercase\expandafter{\romannumeral5} and Section \uppercase\expandafter{\romannumeral6} shows the conclusion.

    \section{Motivation for the RePtA VPP trading strategy}\label{Motivation for the RePtA VPP trading strategy}

\begin{figure}
    \centering
    \includegraphics[width=0.5\textwidth]{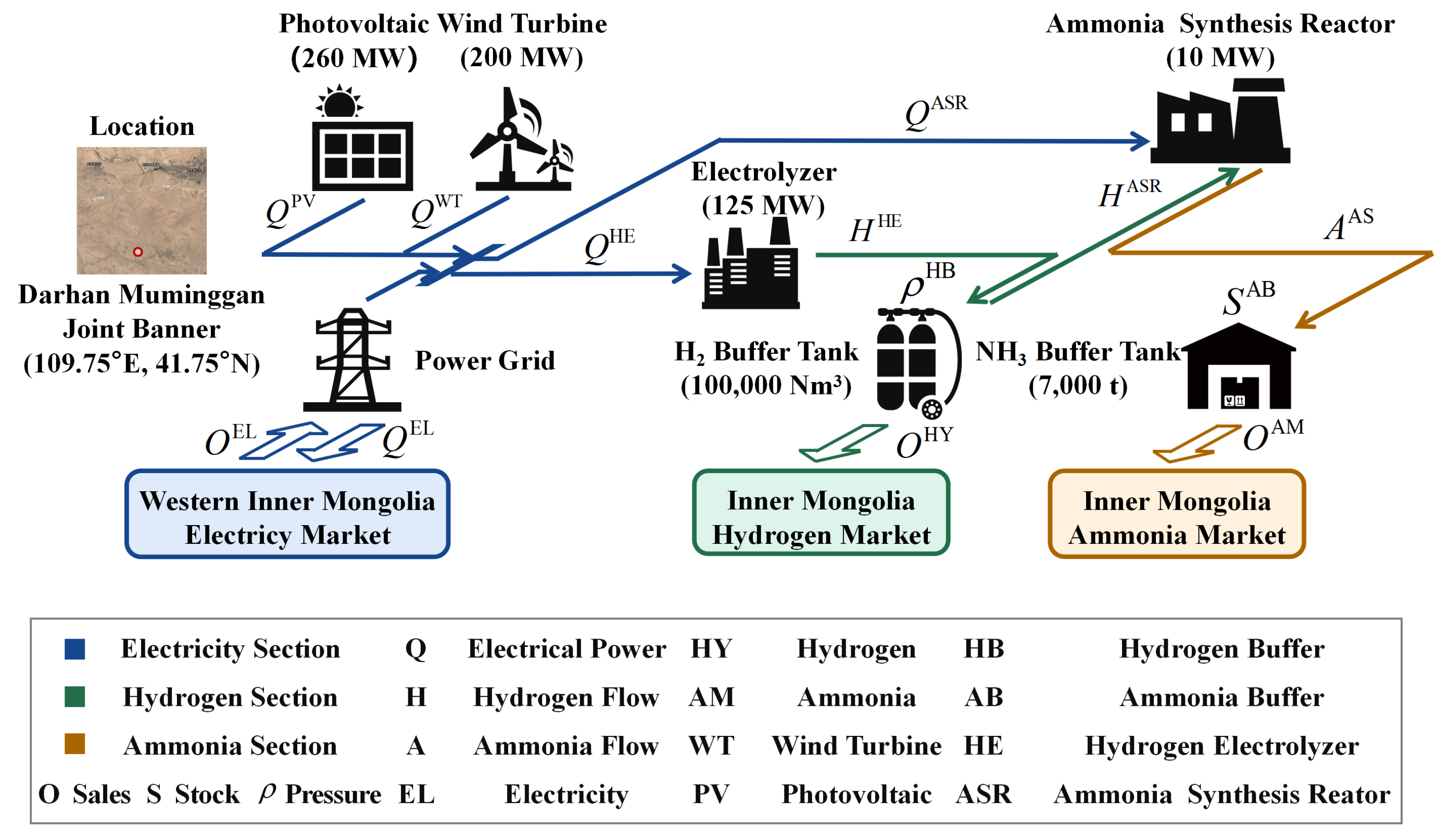}
   \caption{The RePtA VPP project in Inner Mongolia Autonomous Region.}
    \label{The RePtA VPP project in Inner Mongolia Autonomous Region.}
    \end{figure}

    In this section, we present the motivation for the RePtA VPP trading strategy. An actual RePtA VPP project in the Inner Mongolia Autonomous Region is introduced as an example, as presented in Fig. \ref{The RePtA VPP project in Inner Mongolia Autonomous Region.}. 
    
    The RePtA VPP project considered in this study supports the annual production of 100,000 tonnes of ammonia. The markets in which the RePtA VPP participates include the Western Inner Mongolia Electricity Market, Inner Mongolia Hydrogen Market, and Inner Mongolia Ammonia Market.

\begin{figure}
        \centering
        \includegraphics[width=0.45\textwidth]{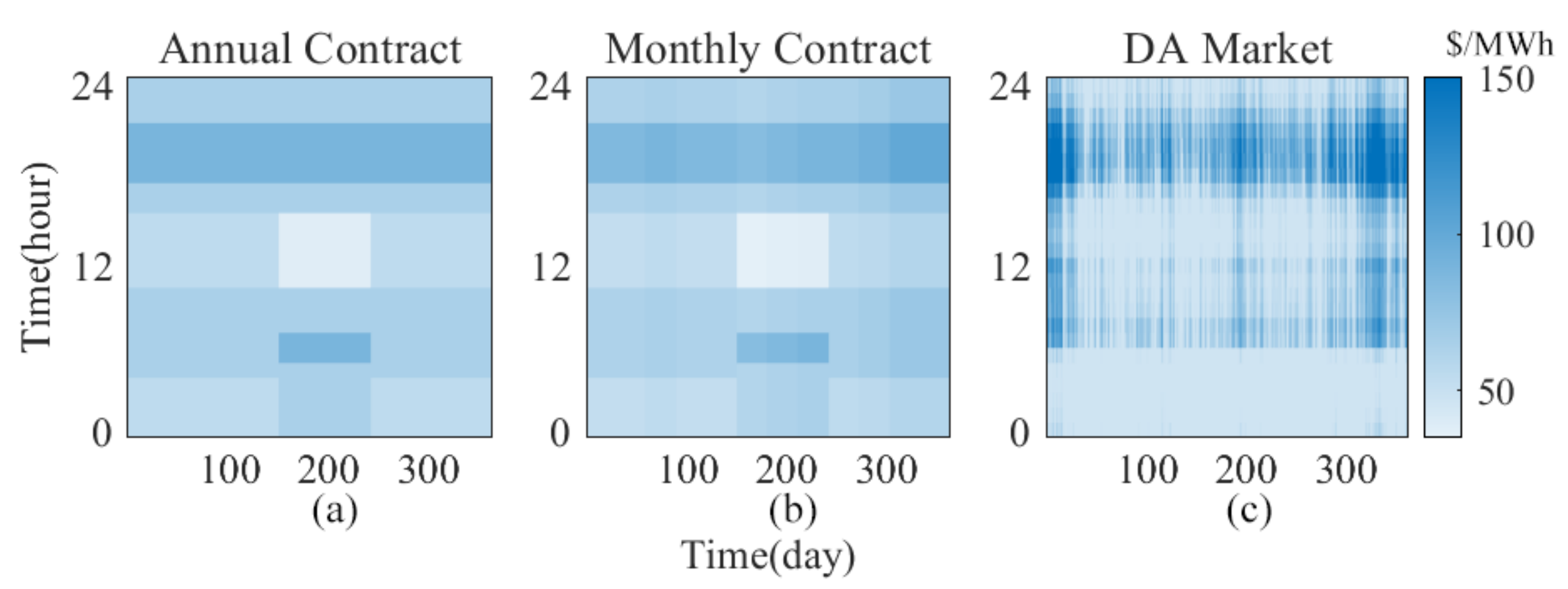}
        \caption{Typical electricity prices in the Western Inner Mongolia Electricity Market. (a) Annual contract. (b) Monthly contract. (c) DA market.}
        \label{Typical electricity prices in the Western Inner Mongolia Electricity Market.}
     \end{figure}

    Fig. \ref{Typical electricity prices in the Western Inner Mongolia Electricity Market.} demonstrates the typical electricity prices in the Western Inner Mongolia Electricity Market. The price of the long-term electricity market follows time-of-use (TOU) pricing. Furthermore, the prices are considerably more variable in the DA market. The price peaks in the evening when there is a high electricity load and a lack of renewables.
 
\begin{figure}
        \centering
        \includegraphics[width=0.45\textwidth]{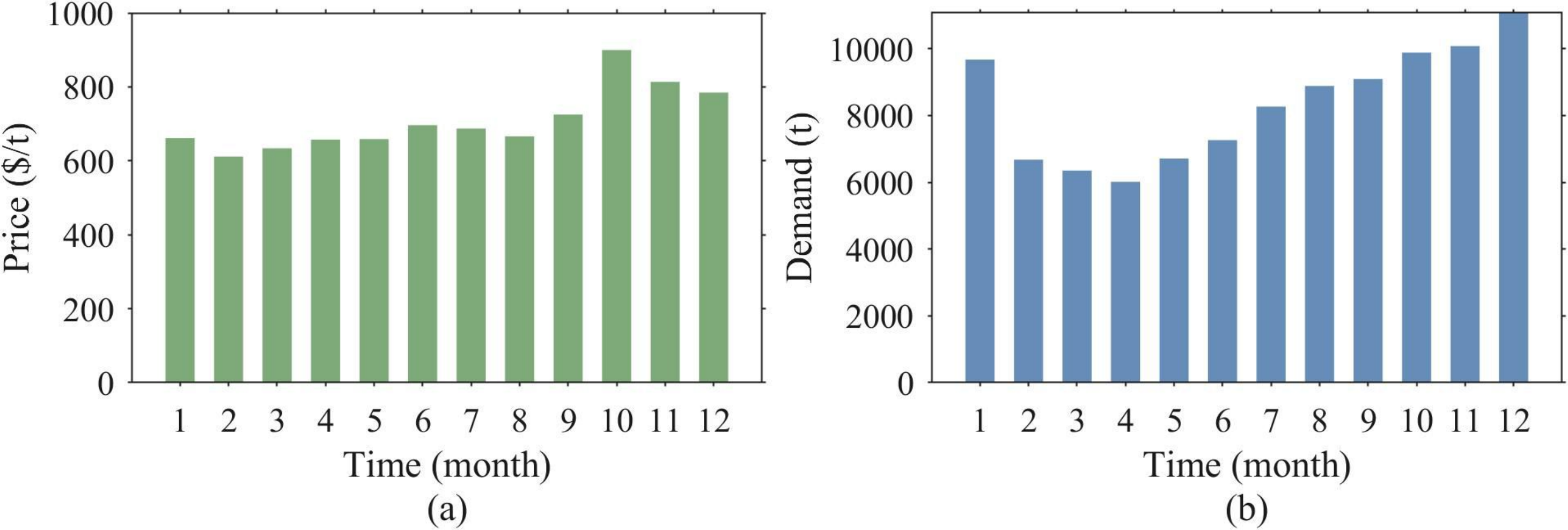}
        \caption{Ammonia price and demand for the RePtA VPP. (a) Ammonia price. (b) Ammonia demand. }
        \label{Ammonia price and demand for the RePtA VPP.}
     \end{figure}

    The historical ammonia price and demand for the RePtA VPP are presented in Fig. \ref{Ammonia price and demand for the RePtA VPP.}. The product demand follows the agricultural cycle, reaching the lowest point in April and peaking in December. Moreover, the RePtA VPP co-produces hydrogen and participates in the Inner Mongolia Hydrogen Market.
    
    \subsection{Power imbalance in the RePtA VPP}
    
    \begin{figure}
    \centering
    \includegraphics[width=0.5\textwidth]{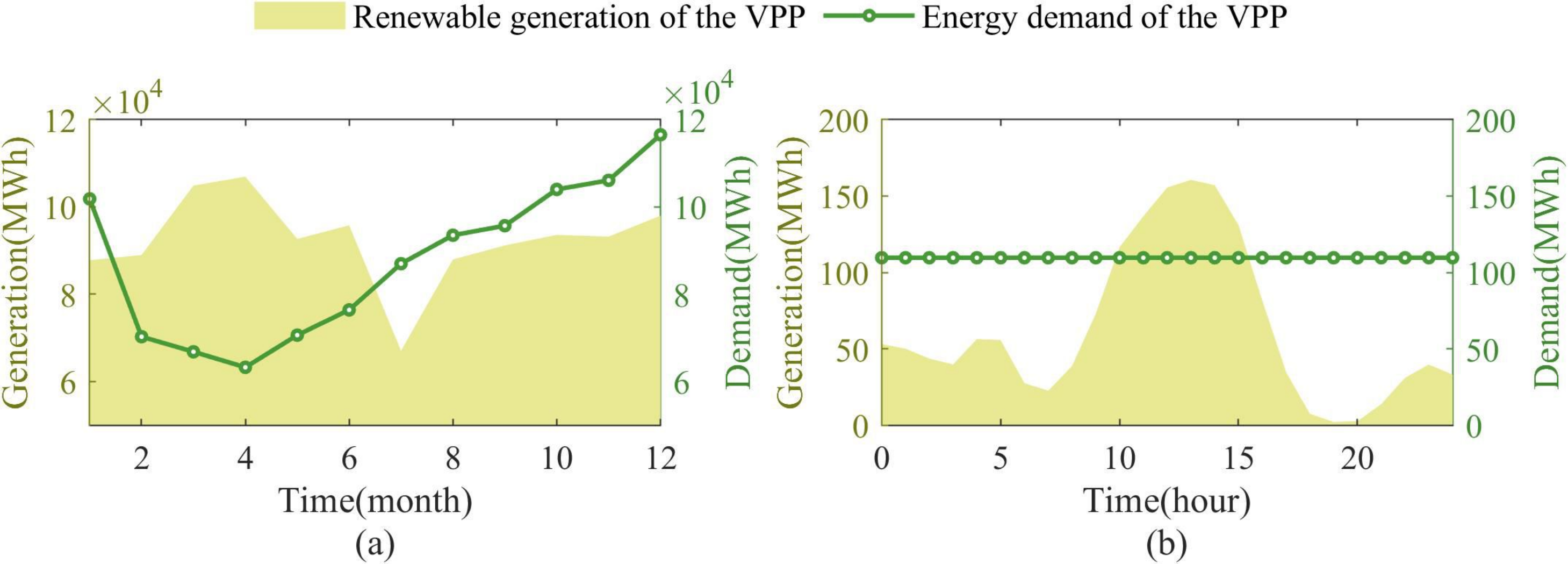}
   \caption{Renewable generation and energy demand of the RePtA VPP. (a) The monthly generation and energy demand. (b) The generation and energy demand on a typical day.}
    \label{VPP generation and energy demand.}
    \end{figure}
    
    According to the conventional operating strategy, the reactor's operation state is changed monthly according to the ammonia demand, which results in power imbalance on multiple timescales. On the monthly timescale, a minor wind period occurs in Inner Mongolia from June to August, as shown in Fig. \ref{VPP generation and energy demand.} (a). The renewable generation of the RePtA VPP in July is usually approximately 74,500 MWh, which is significantly lower than the average output of 102,600 MWh. However, due to agricultural demand, the energy for ammonia production continues to increase during the power shortage period. 

    A similar issue applies to the daily timescale, as shown in Fig. \ref{VPP generation and energy demand.} (b). In a typical day, the load of the RePtA VPP is maintained at 110MW. Nevertheless, renewables have inherent intermittency. The power peaks at 13:00 and falls to 0 MW in the evening, resulting in power shortages.

    \begin{figure}
    \centering
    \includegraphics[width=0.5\textwidth]{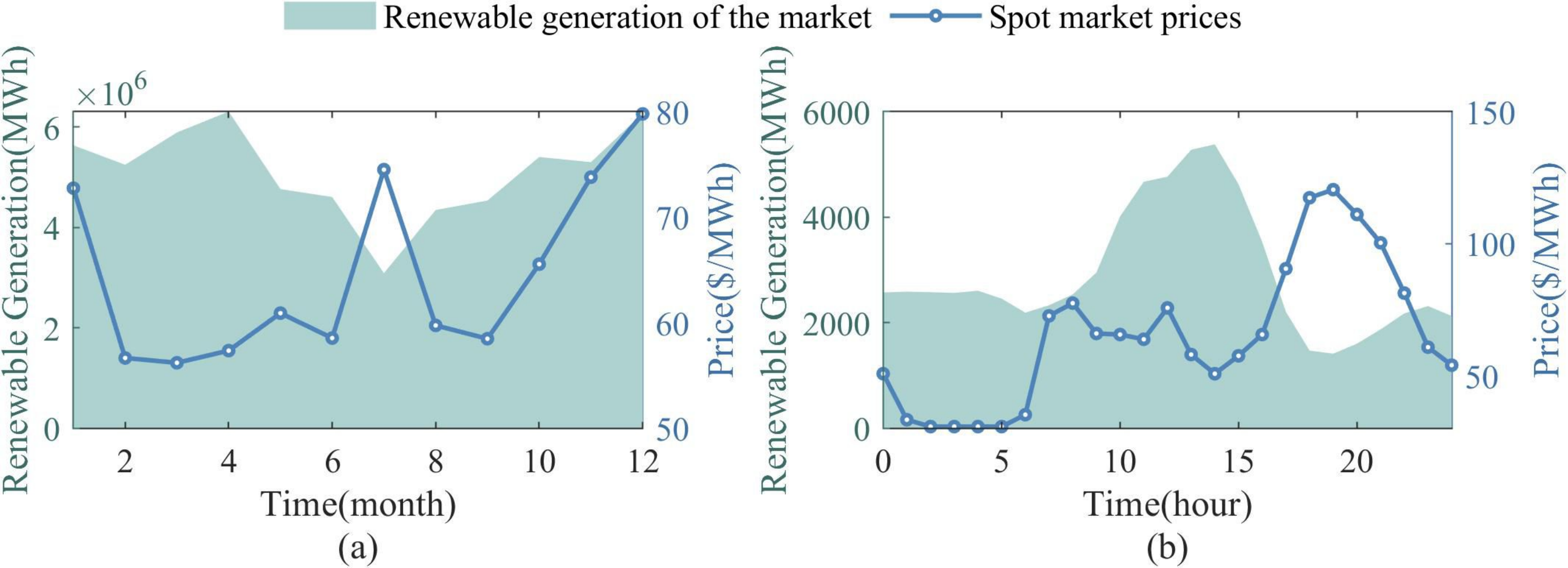}
   \caption{Renewable generation in the market and the corresponding spot market price. (a) The monthly generation in the market and the corresponding spot market price. (b) The generation in the market and the corresponding spot market price on a typical day.}
    \label{Renewable generation in the market and the corresponding spot market price.}
    \end{figure}

    The imbalance can be solved through electricity transactions.

    However, being in the same region, the power profile of the RePtA VPP's units is similar to those in the market. The conventional operating strategy may cause the RePtA VPP to purchase expensive electricity and sell it at a low price. For example, on the monthly timescale, when there is an energy shortage in the RePtA VPP in July, a lack of renewable energy may occur in the market in the same time. The electricity price is \$9.98/MWh higher than the annual average \$64.52/MWh, as presented in Fig. \ref{Renewable generation in the market and the corresponding spot market price.} (a). Furthermore, on the daily timescale, when there is a local energy surplus at 14:00, there is also an abundance of renewable energy on the market. At this time, the price of selling electricity is only \$50.66/MWh, \$15.12/MWh lower than the average daily \$65.78/MWh, as shown in Fig. \ref{Renewable generation in the market and the corresponding spot market price.} (b).

    To sum up, while the electricity market assists the RePtA VPP in achieving power balance, the conventional operating strategy impose the burden of high energy costs.
    
    \subsection{Hydrogen and ammonia buffer systems} 
    
    Hydrogen and ammonia buffer systems include buffer tanks and reactors, which can be found in Fig. \ref{The RePtA VPP project in Inner Mongolia Autonomous Region.}. The buffer tank is capable of storing products for load-shifting. Hydrogen is stored in gas cylinder\cite{c29} and ammonia is stored with low-temperature storage\cite{c30}. 

    The reactor's operation state can be adjusted according to the price fluctuations in the electricity market. The dynamic response of the hydrogen electrolyzer (HE) is pretty fast\cite{c31}. On the hourly time scale, we do not consider the operation constraint of the ramp rate. In contrast, ASR has limited flexibility. It usually takes several hours for a complete working conditions adjustment, which is further explained in Section \ref{Operation and multimarkets trading model}.

\begin{figure}
    \centering
    \includegraphics[width=0.5\textwidth]{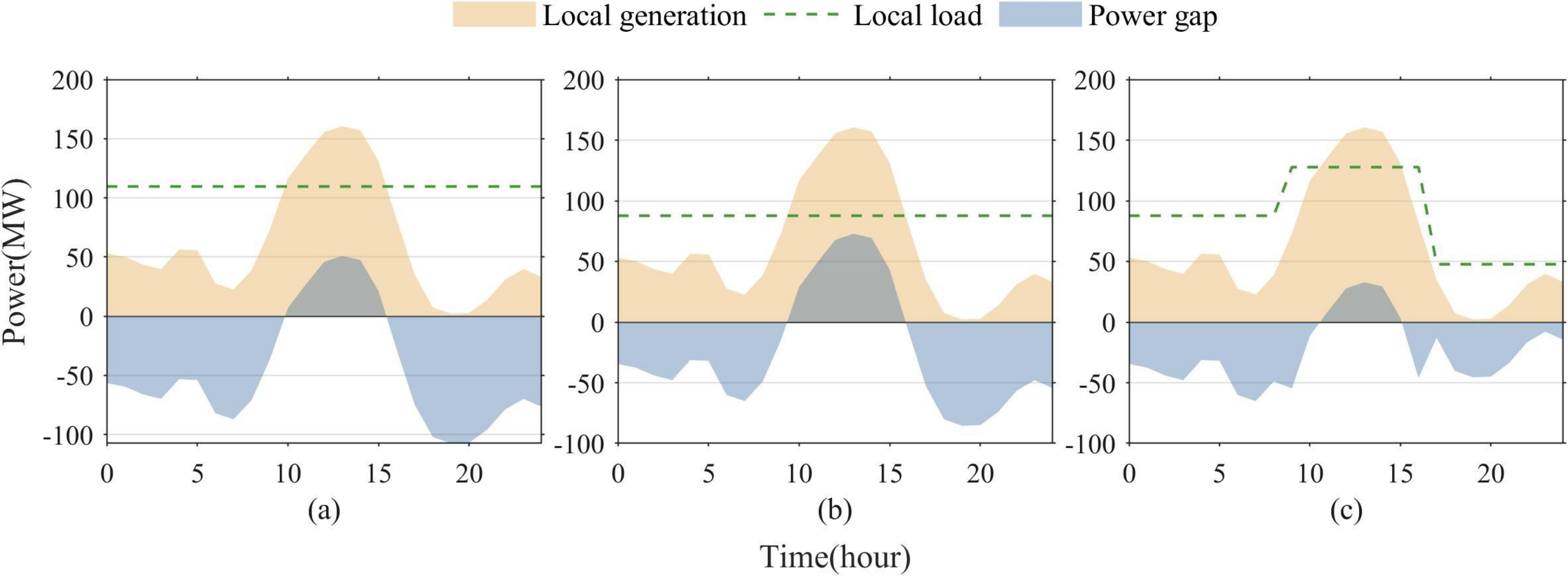}
   \caption{Impact of the hydrogen and ammonia buffer systems on RePtA VPP operation. (a) Conventional operation strategy. (b) With load-shifting via buffer tanks. (c) When considering reactor flexibility. }
    \label{Impact of buffer systems on RePtA VPP operation.}
    \end{figure}

    Fig. \ref{Impact of buffer systems on RePtA VPP operation.}(a) shows the conventional operation strategy, where the power balance entirely relies on transactions with the electricity market. Then, if the hydrogen and ammonia are stored in advance using the buffer tanks, the production load for that period can be reduced, as shown in Fig. \ref{Impact of buffer systems on RePtA VPP operation.} (b). If the reactor's flexibility is further considered, the RePtA VPP can adjust the load to suit the variations in the electricity price, represented in Fig. \ref{Impact of buffer systems on RePtA VPP operation.}(c).
    
    In Section \ref{Motivation for the RePtA VPP trading strategy}, we introduce the motivation for the RePtA VPP trading strategy and the effect of the hydrogen and ammonia buffer systems. The operation and trading model of the RePtA VPP are further discussed thereafter.
    
    \section{RePtA VPP operation and trading model}\label{Operation and multimarkets trading model}
    
    In this section, we introduce the operation model of the RePtA VPP first and then present the market trading models.
    
    \subsection{Operation model}

    We separately establish the balance model of electricity, ammonia, and hydrogen in this subsection. The subscripts $m$, $d$, and $t$ represent monthly, daily, and hourly time scales.

    \subsubsection{Power balance}

    (\ref{power balance}) and (\ref{power to energy}) describe the RePtA VPP's overall power balance. $Q_{m,d,t}^{\rm{EL}}$ and $O_{m,d,t}^{\rm{EL}}$ are the total electricity purchased and sold in electricity market. $P_{m,d,t}^{\rm{WT}}$, $P_{m,d,t}^{\rm{PV}}$ is the power from the wind turbines and photovoltaic plant. Moreover, the power from the HE and ASR are represented by $P_{m,d,t}^{\rm{HE}}$ and $P_{m,d,t}^{\rm{ASR}}$. $\Delta t$ is the unit time interval.
    
    \begin{equation}
    Q_{m,d,t}^{{\rm{WT}}} + Q_{m,d,t}^{{\rm{PV}}} + Q_{m,d,t}^{{\rm{EL}}} - O_{m,d,t}^{{\rm{EL}}} - Q_{m,d,t}^{{\rm{HE}}} - Q_{m,d,t}^{{\rm{ASR}}} = 0
    \label{power balance}
    \end{equation}
    
    \begin{equation}
    Q_{m,d,t}^\tau = P_{m,d,t}^\tau \Delta t,\tau \in \{ {\rm{WT}},{\rm{PV,HE,ASR}}\} 
    \label{power to energy}
    \end{equation}
    
    The power limitation of the ASR is shown in (\ref{ASR power limit}). ${c^{{\rm{ASR}}}}$ is the ASR capacity. Due to catalyst's temperature limits, there are lower and upper limits of $P_{m,d,t}^{{\rm{ASR}}}$\cite{c35}. $\eta_{{\rm{min}}}^{{\rm{ASR}}}$ and $\eta_{{\rm{max}}}^{{\rm{ASR}}}$ are separately set to be 40\% and 100\%.

    \begin{equation}
    \eta ^{\rm{ASR}}_{\rm{min}}{c^{{\rm{ASR}}}} \le P_{m,d,t}^{{\rm{ASR}}} \le \eta ^{\rm{ASR}}_{\rm{max}}{c^{{\rm{ASR}}}}
    \label{ASR power limit}
    \end{equation}
    
    (\ref{Switching of ASR operating status}) considers the transition process of ASR state switching. Adjusting the working conditions of the ASR requires an advance order $S^{ASR}_{m,d,t}$, followed by a series of actions by the operators. Here we use the exponential function to describe this process approximately, as shown in (\ref{Switching of ASR operating status}). The transition time constant ${T_{\rm{con}}^{ASR}}$ is set to be 4 hr . And $T_{{\rm{adj}}}^{{\rm{ASR}}}$ is the adjustment period of ASR.

    \begin{equation}
    \begin{array}{c}
    P_{m,d,t + k}^{ASR} - S_{m,d,t}^{ASR} + \left( {S_{m,d,t + 1}^{ASR} - S_{m,d,t}^{ASR}} \right){\exp ^{ - \frac{k}{{T_{\rm{con}}^{ASR}}}}} = 0\\
    \forall t \in \left\{ {S_{m,d,t + 1}^{ASR} - S_{m,d,t}^{ASR} \ne 0} \right\},k = 1,2, \ldots T_{\rm{adj}}^{ASR}
    \end{array}
    \label{Switching of ASR operating status}
    \end{equation}

   The HE's power is limited by internal operation parameters like current density and temperature\cite{c33}, shown as (\ref{HE power limit}). ${c^{{\rm{HE}}}}$ represents the HE's capacity. $k_{{\rm{min}}}^{{\rm{HE}}}$ and $k_{{\rm{max}}}^{{\rm{HE}}}$ describe the operational bound of the HE, which are separately set to be 5\% and 120\%.

    \begin{equation}
k_{{\rm{min}}}^{{\rm{HE}}}{c^{{\rm{HE}}}} \le P_{m,d,t}^{{\rm{HE}}} \le k_{{\rm{max}}}^{{\rm{HE}}}{c^{{\rm{HE}}}}
    \label{HE power limit}
    \end{equation}

    \subsubsection{Hydrogen balance}
    (\ref{HE power}) - (\ref{HS state5}) represent the hydrogen balance. (\ref{HE power}) models hydrogen production of the electrolyzer, where $H_{m,d,t}^{{\rm{HE}}}$ is the hydrogen production rate and $\eta^{\rm{HE}}$ is the conversion efficiency of the electrolyzer\cite{c34}. And ${\rm{LHV}}_{{\rm{H}}_{2}}$ is the lower heating value of hydrogen.
    
    \begin{equation}
    H_{m,d,t}^{{\rm{HE}}} = \frac{{P_{m,d,t}^{{\rm{HE}}}{\eta ^{{\rm{HE}}}}}}{{\rm{LHV}}_{{\rm{H}}_2}}
     \label{HE power}
    \end{equation}
    
    (\ref{ASR hydrogen consumption}) shows the hydrogen consumption rate ${H_{m,d,t}^{\rm{ASR}}}$ of ammonia synthesis\cite{c35}. $A_{m,d,t}^{{\rm{ASR}}}$ is the production rate of ammonia. $\eta_{{\rm{A2H}}}^{{\rm{ASR}}}$ is 1976 $\rm{NM^3}/\rm{t}$ obtained from the material balance.
    
    \begin{equation}
H_{m,d,t}^{{\rm{ASR}}} = \eta_{{\rm{A2H}}}^{{\rm{ASR}}} A_{m,d,t}^{{\rm{ASR}}}
     \label{ASR hydrogen consumption}
    \end{equation}
    
    (\ref{HS state3}) calculates the hydrogen pressure $\rho _{m,d,t}^{{\rm{HB}}}$ of the hydrogen buffer tank\cite{c36}. $O_{m,d,t}^{\rm{HY}}$ represents the hydrogen sale in the market. ${c^{{\rm{HB}}}}$ is the hydrogen buffer capacity. $R$ is the gas constant. ${T^{{\rm{HB}}}}$ is the mean temperature inside the buffer tank. And ${Mol^{{\rm{HY}}}}$ is the molar mass of hydrogen. (\ref{HS state4}) demonstrates the operational bound of the hydrogen pressure. The final hydrogen pressure is supposed to be equal to the initial value, which is restricted by (\ref{HS state5}).
   
    \begin{equation}
\rho _{m,d,t + 1}^{{\rm{HB}}} = \rho _{m,d,t}^{{\rm{HB}}} + \frac{{R{T^{{\rm{HB}}}}}}{{{c^{{\rm{HB}}}}Mol^{{\rm{HY}}}}}(H_{m,d,t}^{{\rm{HE}}}\Delta t - H_{m,d,t}^{{\rm{ASR}}}\Delta t - O_{m,d,t}^{{\rm{HY}}})
     \label{HS state3}
    \end{equation}
    
    \begin{equation}
\rho _{\min }^{{\rm{HB}}} \le \rho _{m,d,t}^{{\rm{HB}}} \le \rho _{\max }^{{\rm{HB}}}
    \label{HS state4}
    \end{equation}
    
    \begin{equation}
\rho _{{\rm{ini}}}^{{\rm{HB}}} = \rho _{{\rm{end}}}^{{\rm{HB}}}
    \label{HS state5}
    \end{equation}

    \subsubsection{Ammonia balance}
    (\ref{ASR power}) - (\ref{AS state3}) illustrates the ammonia balance. (\ref{ASR power}) models ammonia production\cite{c37}. Conversion coefficient $\eta_{{\rm{P2A}}}^{{\rm{ASR}}}$ is set to be 1.57 $\rm{t}/\rm{MWh}$. 

    \begin{equation}
A_{m,d,t}^{{\rm{ASR}}} = \eta_{{\rm{P2A}}}^{{\rm{ASR}}} P_{m,d,t}^{{\rm{ASR}}}
    \label{ASR power}
    \end{equation}

    The ammonia buffer tank is modeled by (\ref{AS state1}) - (\ref{AS state3}). $S_{m,d,t}^{{\rm{AB}}}$ represents the ammonia stock. $O_{m,d,t}^{\rm{AM}}$ is the ammonia sale in the market. (\ref{AS state2}) limits the amount of ammonia stock, where $c^{\rm{AB}}$ is the ammonia buffer capacity. Similarly, the final ammonia stock is supposed to be equal to the initial value, which is limited by (\ref{AS state3}).
    
    \begin{equation}
    S_{m,d,t}^{{\rm{AB}}} + A_{m,d,t}^{{\rm{ASR}}} \Delta t - O_{m,d,t}^{\rm{AM}} - S_{t + 1}^{{\rm{AB}}} = 0
     \label{AS state1}
    \end{equation}

    \begin{equation}
    0 \le S_{m,d,t}^{{\rm{AB}}} \le c^{\rm{AB}}
    \label{AS state2}
    \end{equation}

    \begin{equation}
S _{{\rm{ini}}}^{{\rm{AB}}} = S_{{\rm{end}}}^{{\rm{AB}}}
     \label{AS state3} 
    \end{equation}
   
    \subsection{Market trading model}
    This subsection models the transactions of the RePtA VPP in the electricity, hydrogen, and ammonia markets. The multi-timescales trading framework is shown in Fig. \ref{Multi-timescales trading framework of the electricity, hydrogen, and ammonia markets.}.
    
    \begin{figure}
        \centering
        \includegraphics[width=0.5\textwidth]{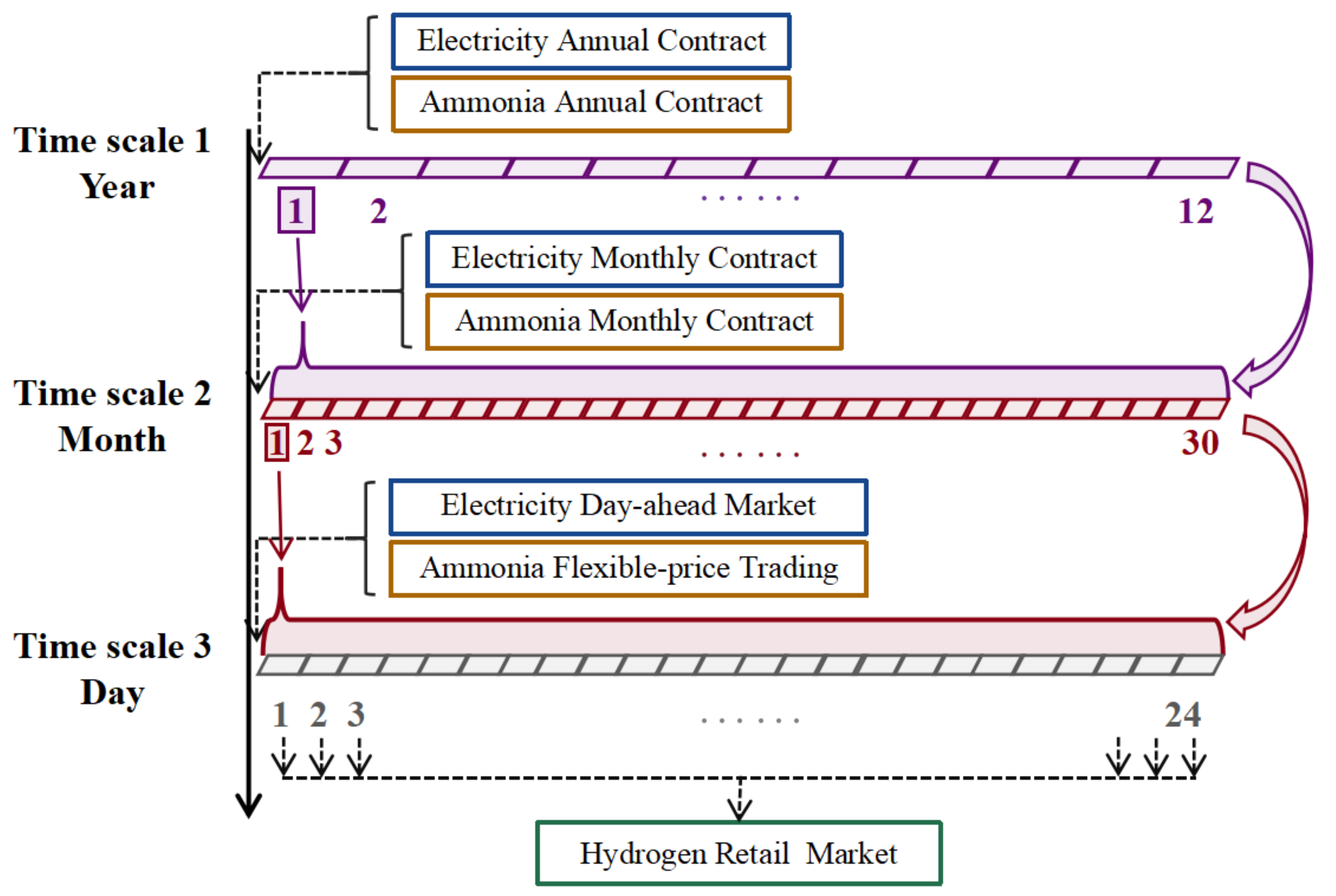}
        \caption{Multi-timescales trading framework of the electricity, hydrogen, and ammonia markets.}
        \label{Multi-timescales trading framework of the electricity, hydrogen, and ammonia markets.}
     \end{figure}
     
    \subsubsection{Electricity trading model}
    
    The two primary markets for electricity are the long-term market (futures market) and the short-term market (spot market). According to local policy, electricity purchases for RePtA VPP can be made in both the long- and short-term markets, while electricity sales can only be made in the short-term market. In this article, we consider the DA market in the spot market and the yearly and monthly contracts in the long-term market. The total amount of electricity bought from the market can be stated as:

    \begin{equation}
    Q_{m,d,t}^{{\rm{EL}}} - (Q_{m,d,t}^{\rm{AC}} + Q_{m,d,t}^{\rm{MC}} + Q_{m,d,t}^{{\rm{DA}}}) = 0
    \label{electricity market}
    \end{equation}    
    
    $ Q_{m,d,t}^{\rm{AC}}$ and $Q_{m,d,t}^{\rm{MC}}$ indicate electricity purchases via annual and monthly contracts. Furthermore, the power purchases and sales in DA market are $Q_{m,d,t}^{{\rm{DA}}}$ and $O_{m,d,t}^{{\rm{EL}}}$, respectively.

    (\ref{On-Grid constrain}) and (\ref{Off-Grid constrain}) represent the limitations for electricity sales and purchases. $\overline{P}^{OFF}$ and $\overline{P}^{ON}$ are the maximum amount of electricity that can be sold or purchased. $s_{m,d,t}$ is a 0-1 variable to prevent the RePtA VPP from simultaneously purchasing and selling electricity. 
 
    \begin{equation}
    0\leq  Q_{m,d,t}^{\rm{EL}} \leq (1- s_{m,d,t})\overline{P}^{OFF}\Delta t
    \label{On-Grid constrain}
    \end{equation}
    
    \begin{equation}
    0\leq  O_{m,d,t}^{\rm{EL}} \leq  s_{m,d,t}\overline{P}^{ON}\Delta t
    \label{Off-Grid constrain}
    \end{equation}

    In an annual contract, the trading curve for a day is identical across the entire month, as shown in (\ref{AC purchase constrain}):
     
     \begin{equation}
Q_{i,d,t}^{{\rm{AC}}} - Q_{i,d + k,t}^{{\rm{AC}}} = 0,k = 1,2,...
    \label{AC purchase constrain}
      \end{equation}
       
    The monthly contract trading follows a similar constraint (\ref{MC purchase constrain}):
       
    \begin{equation}
Q_{i,d,t}^{{\rm{MC}}} - Q_{i,d + k,t}^{{\rm{MC}}} = 0,k = 1,2,...
    \label{MC purchase constrain}
      \end{equation}
  
    In DA markets, electricity trading is more flexible. There are no more constraints on power sales and purchases.

    \subsubsection{Ammonia trading model} 
    For ammonia transactions, there are two modes: fixed-price trading (typically with a long-term contract) and flexible-price trading (usually with temporary customers, and the price is determined by the market supply and demand). The contracts signed for fixed-price trading are conducted as annual and monthly contracts. As for flexible-price trading, the RePtA VPP is supposed to receive orders before the day begins and then make delivery.
    
    The annual ammonia demand from fixed-price trading is $D_{m,d,t}^{\rm{AM,AC}}$ and the monthly ammonia demand from fixed-price trading is $D_{m,d,t}^{\rm{AM,MC}}$. $O_{m,d,t}^{\rm{AM,AC}}$ and $O_{m,d,t}^{\rm{AM,MC}}$ are the corresponding sale, constrained by (\ref{Fixed-price ammonia demand}):
    
    \begin{equation}
    O_{m,d,t}^{{\rm{AM,}}\tau } - D_{m,d,t}^{{\rm{AM,}}\tau } = 0,\tau  \in  \{{\rm{AC,MC}}\} 
    \label{Fixed-price ammonia demand}
    \end{equation}
    
    $D_{m,d,t}^{\rm{AM,DA}}$ is the ammonia demand from flexible-price trading. Furthermore, the RePtA VPP operators will selectively participate in flexible-price trading due to limited ammonia production and long-term contract orders. We assume that the ammonia sold through flexible-price trading is below $d_{m,d,t}^{\rm{AM,DA}}$. Therefore, $O_{m,d,t}^{\rm{AM,DA}}$ satisfies the constraints (\ref{DA Ammonia demand1}) and (\ref{DA Ammonia demand2}):

    \begin{equation}
O_{m,d,t}^{{\rm{AM}},{\rm{DA}}} - D_{m,d,t}^{{\rm{AM}},{\rm{DA}}} \le 0
    \label{DA Ammonia demand1}
    \end{equation}    
    
    \begin{equation}
O_{m,d,t}^{{\rm{AM}},{\rm{DA}}} \le d_{m,d,t}^{{\rm{AM}},{\rm{DA}}}
    \label{DA Ammonia demand2}
    \end{equation}  
    
    The total sales of ammonia $O_{m,d,t}^{\rm{AM}}$ are shown as (\ref{Ammonia sale}):
    
    \begin{equation}
    O_{m,d,t}^{\rm{AM}} = O_{m,d,t}^{\rm{AM,AC}} + O_{m,d,t}^{\rm{AM,MC}} + O_{m,d,t}^{\rm{AM,DA}}
    \label{Ammonia sale}
    \end{equation}

    \subsubsection{Hydrogen trading model} 
    
    Due to the scale restrictions of the wholesale market, we believe that the RePtA VPP can only participate in the daily retail market. $d_{m, d, t}^{{\rm{HY}}}$ is the maximum demand for hydrogen. The hydrogen demand constraint can be expressed as (\ref{hydrogen demand}). 
    
    \begin{equation}
    O_{m,d, t}^{{\rm{HY}}} \le d_{m,d,t}^{\rm{HY}}
    \label{hydrogen demand}
    \end{equation}
    
    In Section \ref{Operation and multimarkets trading model}, the operation and trading models of the RePtA VPP are introduced. Note that the trading cycle of the hydrogen and ammonia markets is strikingly similar to that of the electricity market. With similar sign-up periods, the transaction results for each market can be considered in a timely manner during the rolling optimization. Therefore, it is possible to describe the trading of electricity, ammonia, and hydrogen in a unified framework. Based on this finding, we can extend the available two-stage robust optimization model of the electricity market.

    \section{Multi-timescale, multimarket trading strategy}\label{Multi-timescale, multimarket trading strategy}
    
    In this section, we further determine the multi-timescale, multimarket trading strategy. The receding horizon optimization framework is demonstrated first. Additionally, a two-stage robust optimization model is proposed to solve the RePtA VPP's multi-timescale, multimarket trading problem. C\&CG algorithm is adopted to approach the issue.

    \subsection{Receding horizon optimization framework}

    In this study, the receding horizon optimization is utilized to articulate the multiple timescales\cite{c40}. Fig. \ref{Framework of the proposed the multi-timescale, multimarket trading strategy.} depicts the rolling optimization process. 

    \begin{figure}
        \centering
        \includegraphics[width=0.5\textwidth]{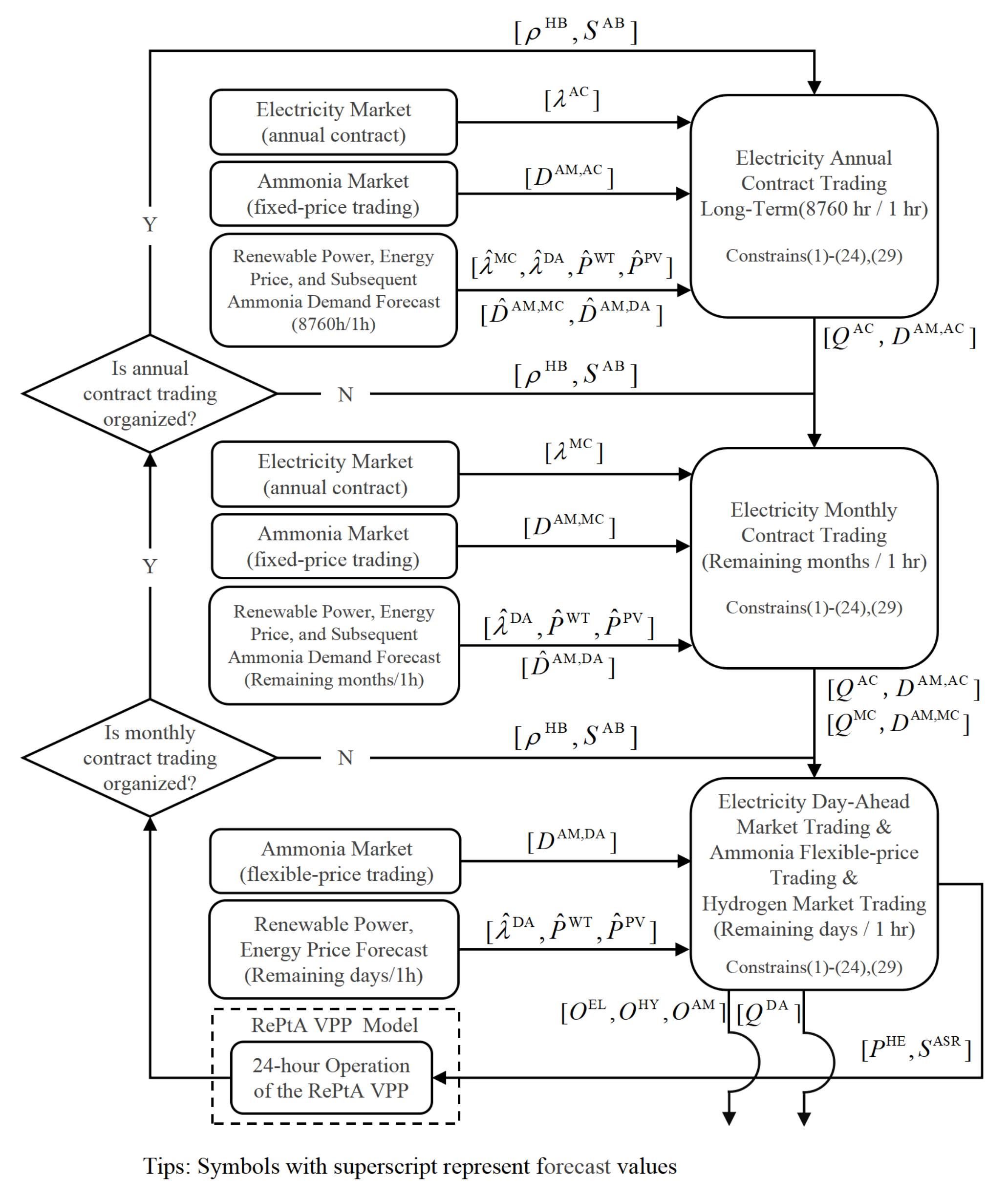}
        \caption{Framework of the proposed the multi-timescale, multimarket trading strategy.}
        \label{Framework of the proposed the multi-timescale, multimarket trading strategy.}
     \end{figure}
    
    The first step is to conduct annual contract trading, which usually takes place before the year begins. The ammonia demand and delivery time are obtained from the signed contracts. Forecast values are used to substitute information that cannot be known explicitly, such as the renewable generation, energy price, and subsequent ammonia demand. Then, the optimization is executed for the first time with an optimization window of 8760 hr and an interval of 1 hr. Based on the results of the solution, the optimal annual contract trading can be determined.
    
    The second step is to conduct monthly contract trading. Take January as an example, the optimization issue is resolved with an optimization window of 8760 hr with an interval of 1 hr. The outcome is used to determine the electricity monthly contract for January.
    
    The following step is to determine spot market trading. For instance, the DA market of January 1 is usually carried out prior to the day, and the RePtA VPP is assumed to participate as a price-taker. The optimized time window is 8760 hours with an interval of 1 hr. The optimization result yields the reference curve for bidding in the spot market on January 1. Then, the above process is repeated, but the optimization window is gradually shortened. For the spot market trading on January 2, the optimal time window is (8760-24) hr.

    \subsection{Two-stage robust optimization model}
    In this subsection, we describe the optimization problem solved in each optimization process. The deterministic model is shown first, and the objective is to maximize the profit within one year, as shown in (\ref{Main objective function}): 
    
    \begin{equation}
    {\rm {Max. } } \left(  R - C_{\rm{mat}} \right)
    \label{Main objective function}
    \end{equation}
    
    subject to 
    
    \begin{equation}
    \rm{Constrain}  (\ref{power balance}) - (\ref{hydrogen demand}).
   \end{equation}
    
    The overall income from product sales is denoted by $R$, including the profit from the sales of electricity, hydrogen, and ammonia. $\mu^{\rm{EL}}$ and $\mu^{{\rm{HY}}}$ are the sales prices of electricity and hydrogen. $\mu^{\rm{AM,AC}}$,$\mu^{\rm{AM,MC}}$, and $\mu^{\rm{AM,DA}}$ are the ammonia price in the annual contract, monthly contract and flexible-price trading.

    \begin{equation}
    \begin{array}{c}
    R = \sum\limits_{m,d,t} {(\mu _{m,d,t}^{{\rm{EL}}}O_{m,d,t}^{{\rm{EL}}} + \mu _{m,d,t}^{{\rm{HY}}}O_{m,d,t}^{{\rm{HY}}}}  + \\
    \mu _{m,d,t}^{{\rm{AM,AC}}}O_{m,d,t}^{{\rm{AM,AC}}} + \mu _{m,d,t}^{{\rm{AM,MC}}}O_{m,d,t}^{{\rm{AM,MC}}} + \mu _{m,d,t}^{{\rm{AM,DA}}}O_{m,d,t}^{{\rm{AM,DA}}})
    \end{array}
    \end{equation}

    $C_{\rm{mat}}$ denotes the cost of raw materials, which refers to the cost of electricity and water. $\lambda^{\rm{AC}}$,$\lambda^{\rm{MC}}$, and $\lambda^{\rm{DA}}$ are the electricity purchase price in the annual contract, monthly contract and DA market. The cost of other auxiliary materials is ignored.
    
    \begin{equation}
    {C_{{\rm{mat}}}} = \Delta t\sum\limits_{m,d,t} {(\lambda _{m,d,t}^{{\rm{AC}}}Q_{m,d,t}^{{\rm{AC}}} + \lambda _{m,d,t}^{{\rm{MC}}}Q_{m,d,t}^{{\rm{MC}}} + \lambda _{m,d,t}^{{\rm{DA}}}Q_{m,d,t}^{{\rm{DA}}})} 
    \end{equation}
    
    Nevertheless, there is some nonnegligible randomness. Then we develop the two-stage robust optimization model that considers stochastic renewable energy. For renewable energy sources, we use confidence bounds and box-like budget constraints to describe ${W} = [P^{\rm{WT}},P^{\rm{PV}}]$\cite{c38}. The uncertainty set can be equivalently expressed as:

    \begin{equation} 
    \Lambda = :\left\{ \begin{array}{c}
    \overline{\chi} _{m,d,t}^{\rm{WT}},\underline{\chi}_{m,d,t}^{\rm{WT}},\overline{\chi} _{m,d,t}^{\rm{PV}},\underline{\chi}_{m,d,t}^{\rm{PV}} \in \{ 0,1\} \\
    P_{m,d,t}^{\rm{WT}} = \hat P_{m,d,t}^{\rm{WT}} + \overline{\chi} _{m,d,t}^{\rm{WT}}\overline{\xi}_{m,d,t}^{\rm{WT}} - \underline{\chi}_{m,d,t}^{\rm{WT}}\underline{\xi} _{m,d,t}^{\rm{WT}}\\
    P_{m,d,t}^{\rm{PV}} = \hat P_{m,d,t}^{\rm{PV}} + \overline{\chi} _{m,d,t}^{\rm{PV}}\overline{\xi}_{m,d,t}^{\rm{PV}} - \underline{\chi}_{m,d,t}^{\rm{PV}}\underline{\xi} _{m,d,t}^{\rm{PV}}\\
    \overline{\chi} _{m,d,t}^{\rm{WT}} + \underline{\chi}_{m,d,t}^{\rm{WT}} \le 1\\
    \overline{\chi} _{m,d,t}^{\rm{PV}} + \underline{\chi}_{m,d,t}^{\rm{PV}} \le 1\\
    \sum\limits_t {(\bar \chi _{m,d,t}^{{\rm{WT}}}  + \overline{\chi}_{m,d,t}^{{\rm{WT}}} )} \le {\rm{K}}\\
    \sum\limits_t {(\bar \chi _{m,d,t}^{{\rm{PV}}} + \bar \chi _{m,d,t}^{{\rm{PV}}})}  \le {\rm{K}}
     \end{array} \right.
    \label{lower-level feasible fields}
    \end{equation}  

    \begin{figure*} 
 	\centering
 	\begin{equation}
    \begin{array}{c}
    \mathop {\max }\limits_{{L_1}} \sum\limits_{m,d,t} {\left( {\mu _{m,d,t}^{{\rm{AM}},{\rm{AC}}}O_{m,d,t}^{{\rm{AM}},{\rm{AC}}} + \mu _{m,d,t}^{{\rm{AM}},{\rm{MC}}}O_{m,d,t}^{{\rm{AM}},{\rm{MC}}} - \lambda _{m,d,t}^{{\rm{AC}}}Q_{m,d,t}^{{\rm{AC}}} - \lambda _{m,d,t}^{{\rm{MC}}}Q_{m,d,t}^{{\rm{MC}}}} \right)} \\
     + \mathop {\min }\limits_{W \in \Lambda } \mathop {\max }\limits_{{L_2}} \sum\limits_{m,d,t} {\left( {\mu _{m,d,t}^{{\rm{EL}}}O_{m,d,t}^{{\rm{EL}}} + \mu _{m,d,t}^{{\rm{HY}}}O_{m,d,t}^{{\rm{HY}}} + \mu _{m,d,t}^{{\rm{AM}},{\rm{DA}}}O_{m,d,t}^{{\rm{AM}},{\rm{DA}}} - \lambda _{m,d,t}^{{\rm{DA}}}Q_{m,d,t}^{{\rm{DA}}}} \right)} 
    \end{array}
    \label{Objective function}
    \end{equation}
    \end{figure*}
    
     $({\hat P_{m,d,t}^{{\rm{WT}}}},{\hat P_{m,d,t}^{{\rm{PV}}}})$ are the forecast of renewable energy power and $(\underline{\xi}_{m,d,t}^{\rm{WT}},\underline{\xi}_{m,d,t}^{\rm{PV}},\overline{\xi} _{m,d,t}^{\rm{WT}},\overline{\xi} _{m,d,t}^{\rm{PV}})$ are the prediction deviation. ${\rm{K}}$ is the uncertainty parameter, which assesses the degree of deviation of renewable energy power. The objective function of the two-stage robust optimization can be formulated as (\ref{Objective function}) with constrains (\ref{power balance}) - (\ref{hydrogen demand}).

    \subsection{Solution methodology}
    A two-stage robust optimization model is proposed in the previous subsection. Nevertheless, given its two-layer structure, the problem has high computational complexity. Benders-style cutting plane methods and the C\&CG algorithm are two common approaches to the exact solution. \cite{c39} shows that the iterations of the C\&CG are lower when the relatively complete recourse assumption holds. Therefore, in this paper, we adopt the C\&CG method to solve the proposed two-stage robust optimization problem.

    In the course of $v = k$ iterations, the objective function of the upper-level problem is:
    
    \begin{equation}
    \begin{array}{c}
    \mathop {\max }\limits_{{L_1}} (\sum\limits_{m,d,t} {(\mu _{m,d,t}^{{\rm{AM,AC}}}O_{m,d,t}^{{\rm{AM,AC}}} + \mu _{m,d,t}^{{\rm{AM,MC}}}O_{m,d,t}^{{\rm{AM,MC}}}} \\
     - \lambda _{m,d,t}^{{\rm{AC}}}Q_{m,d,t}^{{\rm{AC}}} - \lambda _{m,d,t}^{{\rm{MC}}}Q_{m,d,t}^{{\rm{MC}}}) + \theta )
    \end{array}
    \label{upper-level problem relaxed}
    \end{equation}

    \begin{equation}
\begin{array}{c}
\sum\limits_{m,d,t} {(\mu _{m,d,t}^{{\rm{EL}}}O_{m,d,t}^{{\rm{EL}},\nu } + \mu _{m,d,t}^{{\rm{HY}}}O_{m,d,t}^{{\rm{HY}},\nu }} \\
 + \mu _{m,d,t}^{{\rm{AM,DA}}}O_{m,d,t}^{{\rm{AM,DA}},\nu } - \lambda _{m,d,t}^{{\rm{DA}}}Q_{m,d,t}^{{\rm{DA}},\nu })\Delta t \ge \theta 
\end{array}
    \label{upper-level problem - theta}
    \end{equation}

    \begin{equation}
    Q_{m,d,t}^{\rm{AC}} + Q_{m,d,t}^{\rm{MC}}  \ge {f_1}(L_2^\nu )
    \label{upper-level problem - power balance1}
    \end{equation}
    
    \begin{equation}
    {f_1}(L_2^\nu )=O_{m,d,t}^{{\rm{EL}},\nu } + Q_{m,d,t}^{{\rm{HE}},\nu } +  Q_{m,d,t}^{{\rm{ASR}},\nu } - Q_{m,d,t}^{{\rm{DA}},\nu } - Q_{m,d,t}^{{\rm{WT}},\nu } - Q_{m,d,t}^{{\rm{PV}},\nu }
    \label{upper-level problem - power balance2}
    \end{equation}

    \begin{equation}
    \rm{Constrain} (\ref{electricity market})-(\ref{Fixed-price ammonia demand}).
    \end{equation}

    The decision variable of the upper-level problem is ${L_1} = \{ Q_{m,d,t}^{\rm{AC}},Q_{m,d,t}^{\rm{MC}}\}$ and the optimal value of the upper-level function is $F_1^*$. Function (\ref{upper-level problem relaxed}) is the relaxed version of the original objective function (\ref{Objective function}). In (\ref{upper-level problem - theta}), auxiliary variable $\theta$ approximates the value of the lower-level function under the worst-case scenario. $(Q_{m,d,t}^{{\rm{DA}},v'},O_{m,d,t}^{{\rm{EL}},\nu},O_{m,d,t}^{{\rm{HY}},\nu},P_{m,d,t}^{{\rm{HE}},\nu},P_{m,d,t}^{{\rm{ASR}},\nu},P_{m,d,t}^{{\rm{WT,}}\nu} , P_{m,d,t}^{{\rm{PV,}}\nu})$ are the optimal values obtained by the lower-level problem at iteration $\nu = k-1 $. (\ref{upper-level problem - power balance1}) and (\ref{upper-level problem - power balance2}) represent the power constraints in upper-level problem. 

    In the course of $v = k$ iterations, the objective function of the lower-level problem is:

    \begin{equation}
    \begin{array}{c}
    \mathop {\min }\limits_{W \in \Lambda } \mathop {\max }\limits_{{L_2}} \sum\limits_{m,d,t} {(\mu _{m,d,t}^{{\rm{EL}}}O_{m,d,t}^{{\rm{EL}}} + \mu _{m,d,t}^{{\rm{HY}}}O_{m,d,t}^{{\rm{HY}}}} \\
     + \mu _{m,d,t}^{{\rm{AM,DA}}}O_{m,d,t}^{{\rm{AM,DA}}} - \lambda _{m,d,t}^{{\rm{DA}}}Q_{m,d,t}^{{\rm{DA}}})
    \end{array}
    \label{lower-level problem}
    \end{equation}

    \begin{equation}
Q_{m,d,t}^{{\rm{WT}}} + Q_{m,d,t}^{{\rm{PV}}} + Q_{m,d,t}^{{\rm{DA}}} - O_{m,d,t}^{{\rm{EL}}} - Q_{m,d,t}^{{\rm{HE}}} - Q_{m,d,t}^{{\rm{ASR}}} \ge {f_2}(L_1^\nu )
    \label{lower-level problem - power balance1}
    \end{equation}

    \begin{equation}
    {f_2}(L_1^\nu )= - Q_{m,d,t}^{\rm{AC},\nu } - Q_{m,d,t}^{\rm{MC},\nu }
    \label{lower-level problem - power balance2}
    \end{equation}    
    
    \begin{equation}
    \rm{Constrain}  (\ref{power balance})-(\ref{AS state3}),(\ref{DA Ammonia demand1})-(\ref{hydrogen demand}),(\ref{lower-level feasible fields}).
   \end{equation}

    The decision variable of the lower-level problem is ${L_2} = (Q_{m,d,t}^{{\rm{DA}}},O_{m,d,t}^{{\rm{EL}}},O_{m,d,t}^{{\rm{HY}}},P_{m,d,t}^{{\rm{HE}}},P_{m,d,t}^{{\rm{ASR}}})$ and the optimal value of the lower-level function is $F_2^*$. $( Q_{m,d,t}^{{\rm{AC,}}\nu }, Q_{m,d,t}^{{\rm{MC,}}\nu })$ are the optimal values obtained by of the upper-level problem at iteration $\nu = k $. (\ref{lower-level problem - power balance1}) and (\ref{lower-level problem - power balance2}) represent the power constraints in lower-level problem.    
    (\ref{lower-level problem}) is a bi-level model, where the inner-layer minimization is a linear problem. According to the strong duality theorem, the inner-layer minimization can be transformed into maximization form. Then, the solution to the primary problem can be found by solving the dual issue. 
    
    In Section \ref{Multi-timescale, multimarket trading strategy}, we present the decision-making process of the multimarket trading strategy. The case studies are shown in the following.

    \section{Case Studies}\label{Case Studies}
    
    In this section, the numerical simulation based on an actual project in Inner Mongolia Autonomous Region is introduced. In addition, the results of a sensitivity analysis regarding the hydrogen and ammonia buffering systems are shown.
    
    \subsection{Case setup}

    The LCOA is adopted as the evaluation index to evaluate the production cost, which is calculated by (\ref{LCOA1}). Table \ref{The major investment parameters.} shows the major investment parameters of the RePtA VPP.
     
    \begin{table}[h]
    \centering
    \caption{The major investment parameters.}
    \begin{tabular}{ c c c }\hline\hline
Symbol      &  Parameter                             & Value   \\\hline
$c^{{\rm{WT}}} $ & Wind turbine capacity (MW) & 200\\
$c^{{\rm{PV}}} $ & Photovoltaic plant capacity (MW) & 260\\
$c^{\rm{HB}}$    & Hydrogen buffer capacity ($\rm{Nm^3}$) & 100,000 \\
$c^{{\rm{AB}}}$  & Ammonia buffer capacity (t) & 7,000\\
${c^{{\rm{HE}}}}$& HE capacity (MW)& 125 \\
$c^{\rm{ASR}}$   & ASR capacity (MW) & 10 \\
$\gamma ^{\rm{WT}}$    &Wind turbine cost (\$/MW)  & 696,800        \\
$\gamma ^{\rm{PV}}$    &Photovoltaic unit cost (\$/MW)  & 556,200        \\
$\gamma ^{\rm{HB}}$    &Hydrogen buffer tank cost (\$/$\rm{Nm^3}$)  & 37.33   \\
$\gamma ^{\rm{AB}}$    &Ammonia buffer tank cost (\$/t)     & 504.12      \\
$\gamma ^{\rm{HE}}$    &HE cost (\$/MW)     & 447,900      \\
${C^{{\rm{ASR}}}}$    &ASR cost (\$/MW)     & 49,269,000   \\$r$  & Interest rate & $6\%$ \\
    \hline\hline
    \end{tabular}
    \label{The major investment parameters.}
    \end{table}

    \begin{figure*} 
 	\centering
    \begin{equation}
{\rm{LCOA = }}({C_{{\rm{inv}}}}{\rm{ + }}{C_{{\rm{OM}}}}{\rm{ + }}{C_{{\rm{mat}}}} - \sum\limits_{m,d,t} {(\mu _{m,d,t}^{{\rm{EL}}}O_{m,d,t}^{{\rm{EL}}}{\rm{ + }}\mu _{m,d,t}^{{\rm{HY}}}O_{m,d,t}^{{\rm{HY}}})){\rm{/}}\sum\limits_{m,d,t} {O_{m,d,t}^{{\rm{AM}}}} } 
    \label{LCOA1}
    \end{equation}
    \end{figure*}

    \begin{figure*} 
 	\centering
    \begin{equation}
{C_{{\rm{inv}}}} = \frac{{r{{(1 + r)}^n}}}{{{{(1 + r)}^{n - 1}}}}\left( {{\gamma ^{{\rm{WT}}}}{c^{{\rm{WT}}}} + {\gamma ^{{\rm{PV}}}}{c^{{\rm{PV}}}} + {\gamma ^{{\rm{HB}}}}{c^{{\rm{HB}}}} + {\gamma ^{{\rm{AB}}}}{c^{{\rm{AB}}}} + {\gamma ^{{\rm{HE}}}}{c^{{\rm{HE}}}} + {C^{{\rm{ASR}}}}} \right)
\label{Investment costs}
    \end{equation}
    \end{figure*} 
    
    $C_{\rm{inv}}$ and $C_{\rm{OM} }$ indicate the investment and O\&M costs, calculated by (\ref{Investment costs}) and (\ref{O&M costs}). $\eta$ is the ratio of annual O\&M costs to investment costs, set to be 3\% in this study. $\gamma ^{\rm{WT}}$, ${\gamma ^{\rm{PV}}}$, $\gamma ^{\rm{HE}}$, ${\gamma ^{\rm{HB}}}$, and ${\gamma ^{\rm{AB}}}$ are the unit investment costs of various facilities. ${c^{\rm{WT}}}$, ${c^{\rm{PV}}}$, ${c^{\rm{HE}}}$,  ${c^{\rm{HB}}}$, and ${c^{\rm{AB}}}$ are the corresponding equipment capacities. And ${C^{{\rm{ASR}}}}$ represents cost of ASR. 
    
    \begin{equation}
    C_{\rm{OM}} = \eta C_{\rm{inv}}
    \label{O&M costs}
    \end{equation}

    \subsection{Simulation results}

    \begin{figure}
        \centering
        \includegraphics[width=0.5\textwidth]{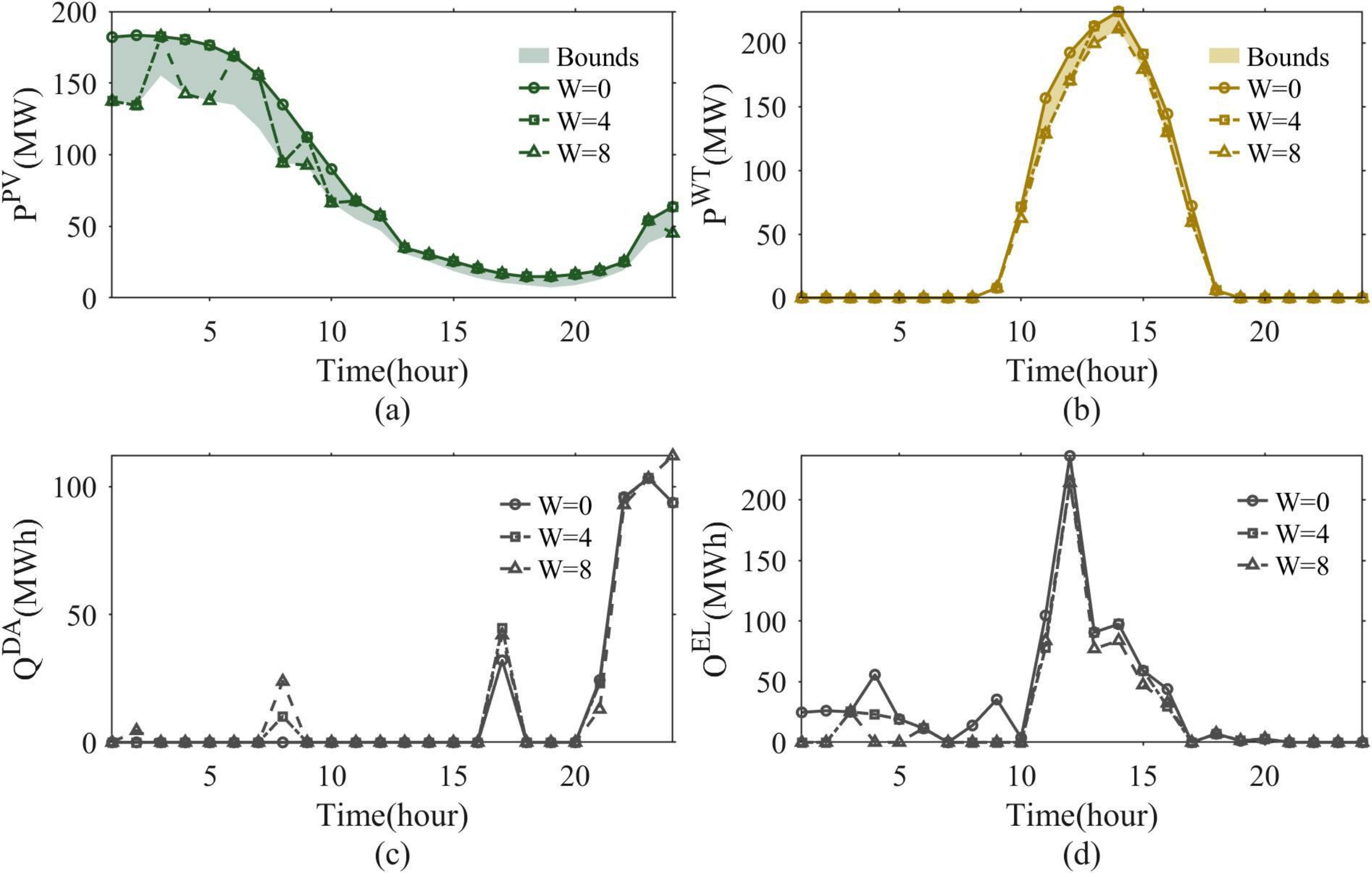}
        \caption{Optimization results on January 1 with different $K$. (a) Wind turbine generation. (b) Photovoltaic plant generation. (c) Power purchased in the spot market. (d) Power sold in the spot market.}
        \label{Robust optimization results on January 1.}
     \end{figure}
     
    Fig. \ref{Robust optimization results on January 1.} depicts the RePtA VPP operation in January 1 with different $K$. The larger $K$ is, the more electricity the RePtA VPP buys and the less electricity it sells. In this paper, we set the value of K to 4.

    \begin{figure*}
        \centering
        \includegraphics[width=0.98\textwidth]{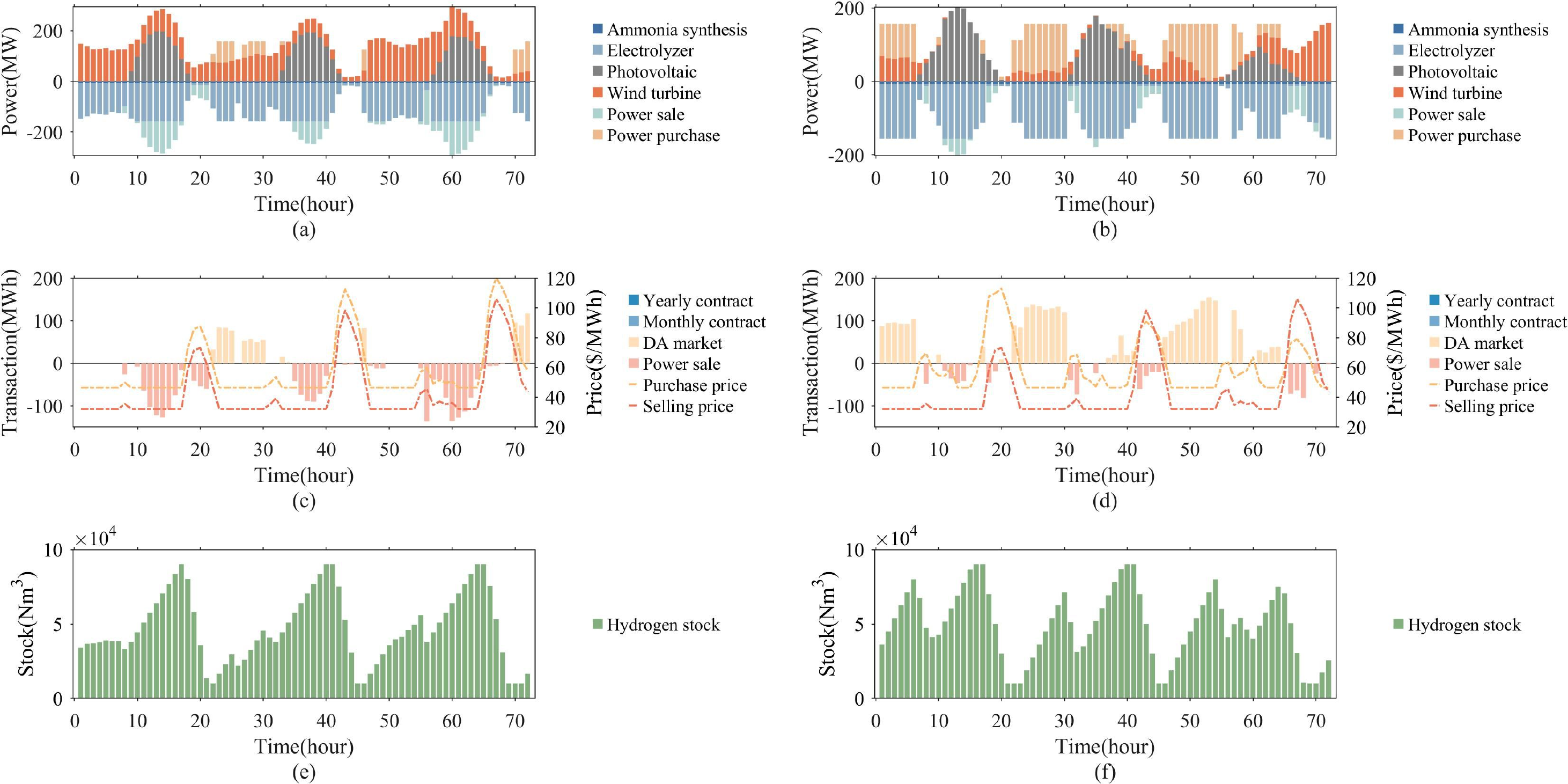}
        \caption{Typical operation of the RePtA VPP in winter and summer. (a) Power balance of the RePtA VPP from February 15 - 18. (b) Power balance of the RePtA VPP from August 15 - 18. (c) Electricity transaction from February 15 - 18. (d) Electricity transaction from August 15 - 18. (e) Hydrogen stock from February 15 - 18. (f) Hydrogen stock from August 15 - 18.}
        \label{The RePtA VPP operation.}
     \end{figure*}

    Furthermore, we choose two typical periods (February 15 – 18 and August 15 - 18 ) to demonstrate how the RePtA VPP functions in winter and summer. Inner Mongolia has high and low mean wind speeds in these two periods, respectively. The red dashed line in Fig. \ref{The RePtA VPP operation.} (c) represents the fluctuating electricity price from 15 to 18 February. Under the pricing incentives, the RePtA VPP employs buffer systems for arbitrage. For example, approximately the 20th hour, the RePtA VPP reduces the electrolyzer power from 150MW to 6.25MW, and the less flexible ASR is held at 7.6 MW. Although ammonia synthesis still consumes approximately 22,000 Nm$^3$ of hydrogen per hour, the hydrogen comes from the buffer tanks, as presented in Fig. \ref{The RePtA VPP operation.} (e).

    Fig. \ref{The RePtA VPP operation.} (b), (d), and (f) depict the RePtA VPP operation from 15 to 18 August. The transactions in the electricity market follow a typical trading pattern, as shown in Fig. \ref{The RePtA VPP operation.} (d). The RePtA VPP purchases a sizable amount of electricity every day between 0:00 - 6:00 and 22:00 - 24:00 when the electricity price is low. Such cyclicality is also reflected in the hydrogen stock. In Fig. \ref{The RePtA VPP operation.} (f), the hydrogen stock peaks at approximately 6:00 and 16:00 every day and is consumed in the morning and evening.

    On longer time scales, the ammonia buffer exhibits a significant load-shifting effect. Fig. \ref{Production,sale and storage of ammonia throughout the year.} shows the production, demand, and stock of ammonia. The high of the ammonia stock is above 4,000 t before June and November when the electricity price peaks in Fig. \ref{Renewable generation in the market and the corresponding spot market price.} (c). Using the large-scale ammonia buffer tank, the RePtA VPP implements inter-week storage to cope with limited wind resources in the summer and high electricity rates in the winter.
     
    \begin{figure}
        \centering
        \includegraphics[width=0.48\textwidth]{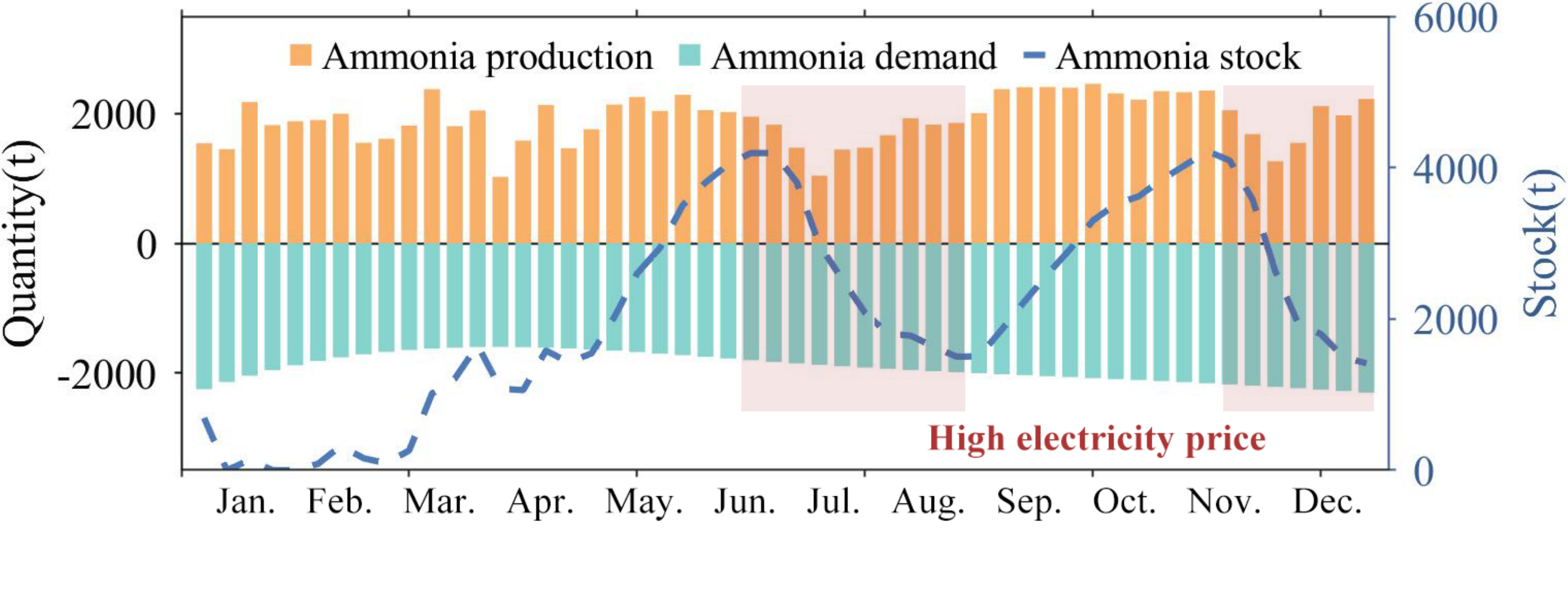}
        \caption{Production,sale and storage of ammonia throughout the year.}
        \label{Production,sale and storage of ammonia throughout the year.}
     \end{figure}

    \subsubsection{Sensitivity analysis of electricity market trading modes}
    In this subsection, we discuss the LCOA in the case of different electricity trading modes. Three modes are compared: participation in both spot and long-term markets, participation in the spot market only, and adopt TOU pricing. The TOU price is assumed to be numerically equal to the annual contract price. Additionally, the trading for hydrogen and ammonia remain unchanged. 

    The results are shown in Table \ref{LCOA under different electricity trading modes}. Participating in the long-term and spot electricity markets or the spot market only results in an LCOA of \$222.73/t. Due to the abundance of renewable energy sources, locking-in power from the long-term market may result in energy waste. Therefore, the operator of the RePtA is unwilling to participate in the long-term trading, which is in line with the situation shown in Fig. \ref{The RePtA VPP operation.} (c) (d). However, if TOU pricing is adopted, the LCOA rises significantly to \$405.38/t.
    
    \begin{table}
    \centering
    \caption{LCOA under different electricity trading modes}
    \label{LCOA under different electricity trading modes}
    \begin{tabular}{cccc}
    \hline\hline
     & Long-term\& Spot market & Spot market  & TOU\\
     \hline
    LCOA($\rm{\$/t)}$ & 222.73 & 222.73 & 405.38  \\\hline\hline
    \end{tabular}
    \end{table}

\begin{figure}
        \centering
        \includegraphics[width=0.45\textwidth]{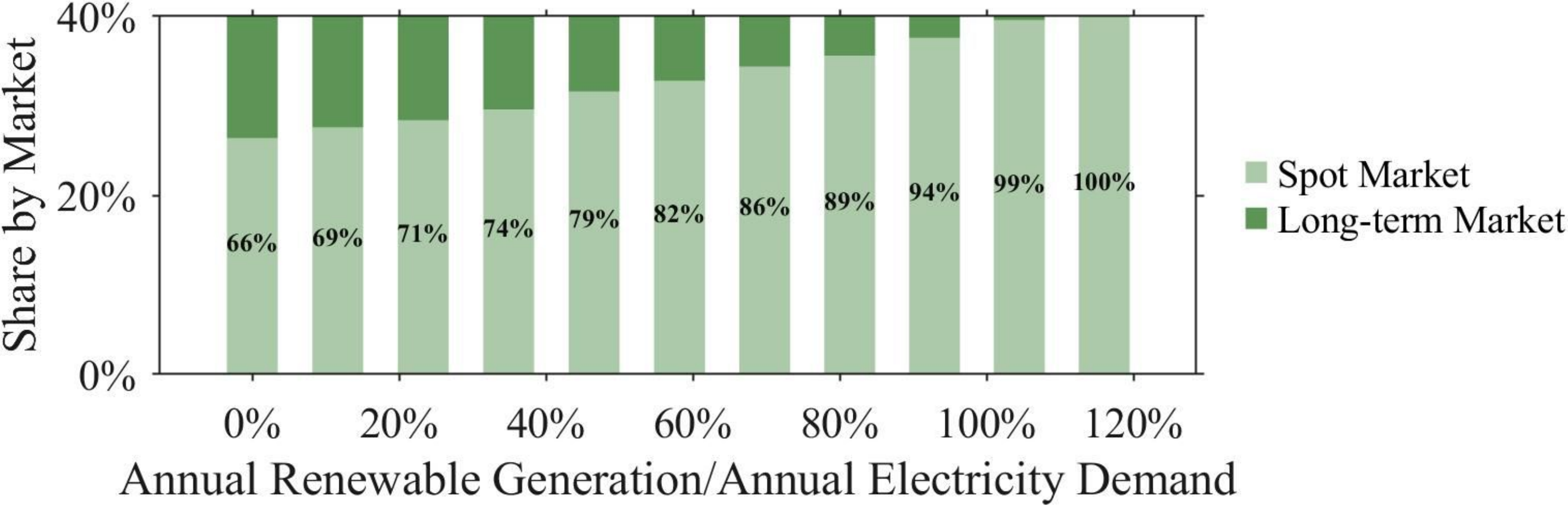}
        \caption{Share of electricity purchases in different markets.}
        \label{Share of electricity purchases in different markets.}
     \end{figure}

    Furthermore, we consider the optimal electricity trading strategy for the RePtA VPP with different generation capacities. Fig. \ref{Share of electricity purchases in different markets.} depicts the share of electricity purchases in different markets. The RePtA VPP purchases approximately 34\% of the electricity from the long-term market when it does not have any local units. Then when the annual renewable generation reaches nearly 1.2 times the annual demand, the RePtA VPP no longer purchases electricity from the long-term market. It prefers to trade in the flexible short-term market in this case.

    \subsubsection{Sensitivity analysis of hydrogen and ammonia buffer capacity}
    
\begin{figure}
        \centering
        \includegraphics[width=0.5\textwidth]{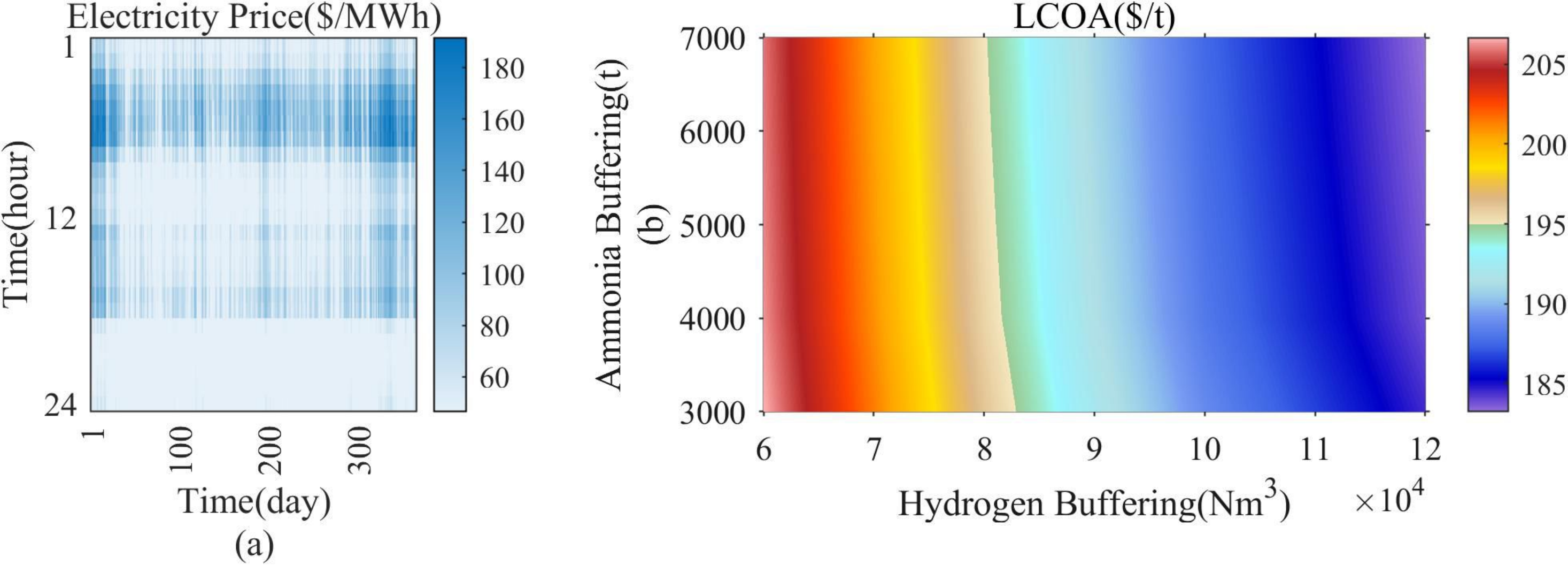}
        \caption{Sensitivity analysis with hourly fluctuations in prices. (a) Market price adopted. (b) Calculation results of LCOA.}
        \label{Sensitivity analysis with hourly fluctuations in prices.}
    \end{figure}

\begin{figure}
        \centering
        \includegraphics[width=0.5\textwidth]{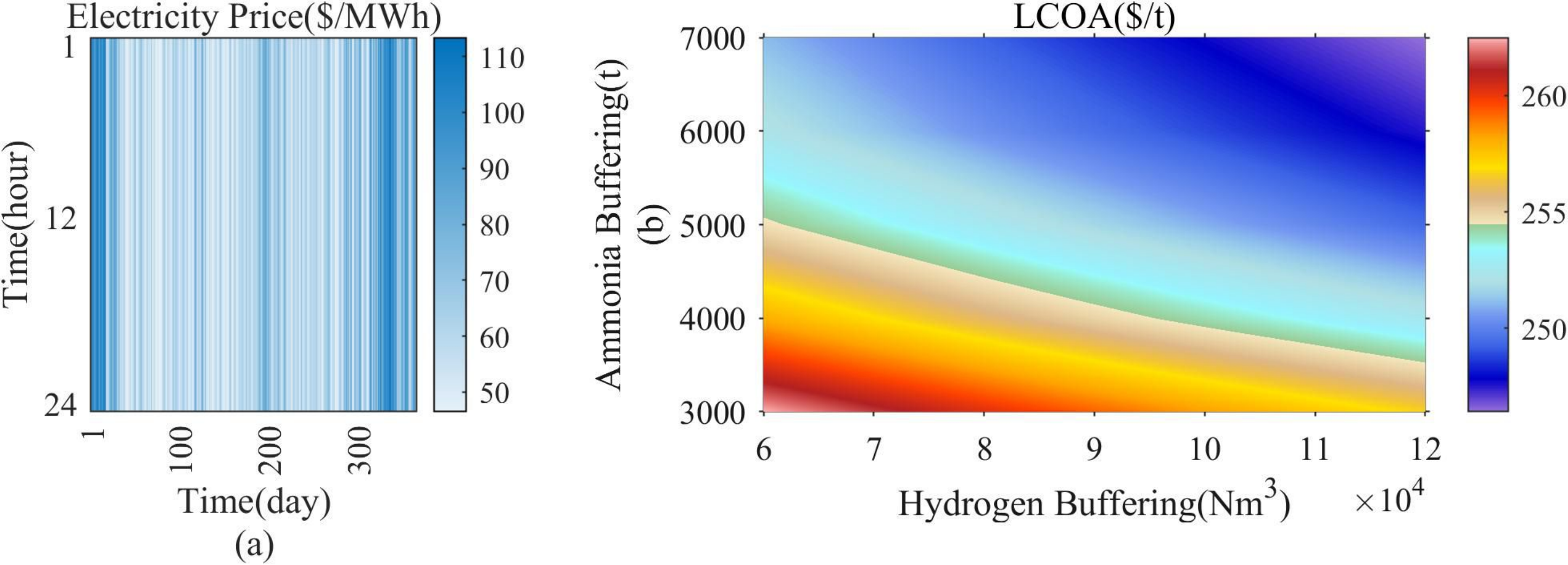}
        \caption{Sensitivity analysis with daily fluctuations in prices. (a) Market price adopted. (b) Calculation results of LCOA.}
        \label{Sensitivity analysis with daily fluctuations in prices.}
    \end{figure}

    In this subsection, we conduct a sensitivity analysis demonstrating the effect of hydrogen and ammonia buffer hedging against different price fluctuations. In the first case, only hourly changes are maintained, as depicted in \ref{Sensitivity analysis with hourly fluctuations in prices.} (a). In the second case, only the day-level price fluctuations are in reserve, as presented in \ref{Sensitivity analysis with daily fluctuations in prices.} (a).

    Fig. \ref{Sensitivity analysis with hourly fluctuations in prices.} shows the LCOA when there are only hourly fluctuation. Unexpectedly, ammonia storage does not to appear to affect energy costs within the simulation scope. The LCOA rises by \$1/t when the ammonia buffer capacity increases from 3,000 to 7,000 Nm$^3$. In contrast, the hydrogen buffer has a significant effect on reducing LCOA. Increasing hydrogen buffer capacity from 6,000 to 12,000 Nm$^3$ reduces LCOA by approximately \$22/t. Considering that the ASR consumes hydrogen at a rate of 2,200 Nm$^3$/hr, the hydrogen buffer tank should ideally be able to store approximately 6 hours of hydrogen for ammonia production.
        
    Fig. \ref{Sensitivity analysis with daily fluctuations in prices.} illustrates the simulation results when electricity prices have daily differences. Both the hydrogen and ammonia buffers help to reduce costs, while the ammonia buffer has a more notable impact. In the tested range, increasing the ammonia buffer from 3,000 t to 7,000 t yields an LCOA reduction of \$11/t, while increasing the hydrogen buffer from 6,000 Nm$^3$ to 12,000 Nm$^3$ lowers the LCOA by approximately \$5/t.
    
    Comparing Fig. \ref{Sensitivity analysis with hourly fluctuations in prices.} and Fig. \ref{Sensitivity analysis with daily fluctuations in prices.} reveals that the hydrogen buffer is more effective for hourly price fluctuations, while the ammonia buffer is practical for longer-term price fluctuations. This is consistent with the findings in \cite{c41}\cite{c42}.

    \subsubsection{Sensitivity analysis of reactor flexibility}
\begin{figure}
        \centering
        \includegraphics[width=0.5\textwidth]{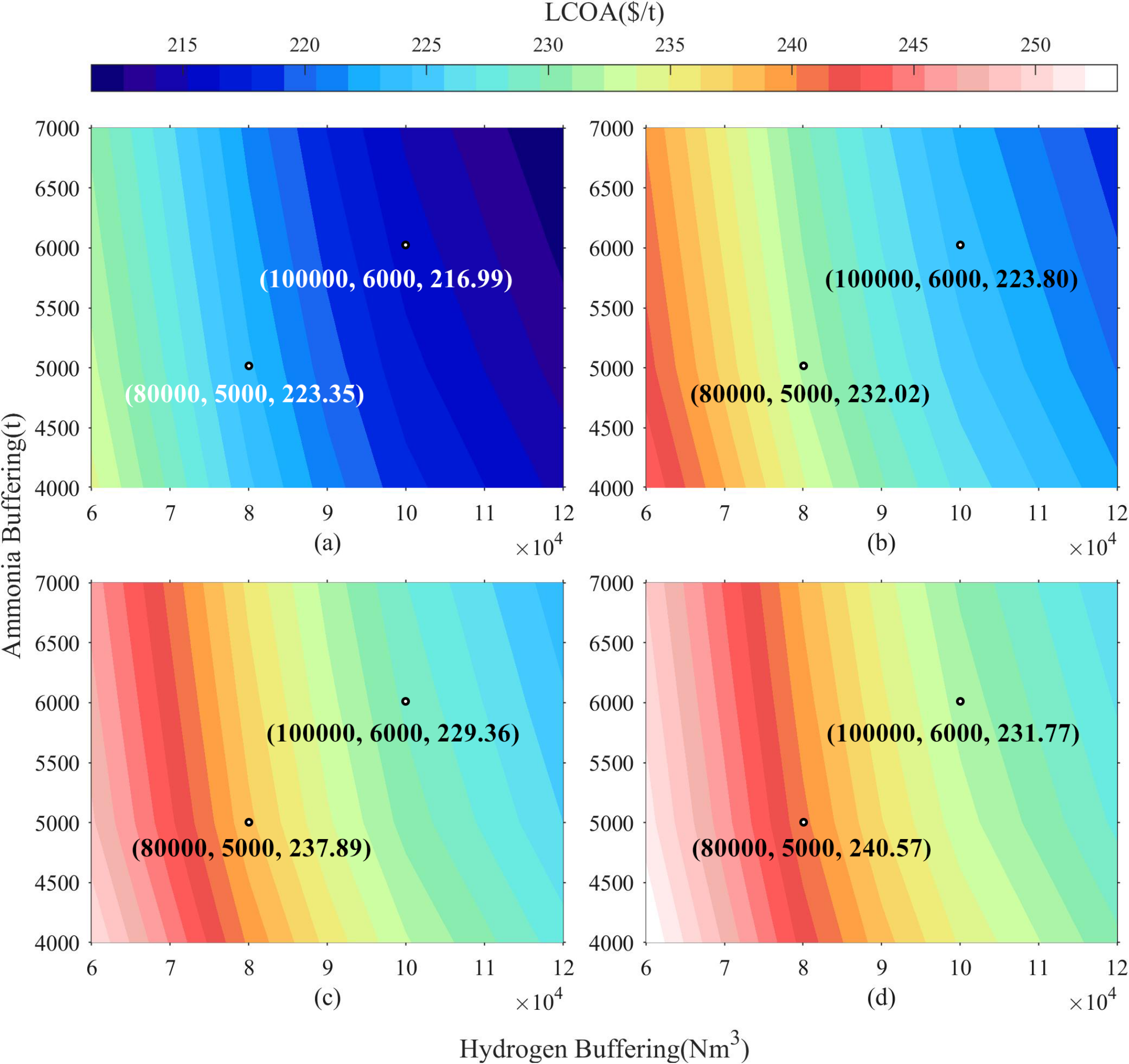}
        \caption{Impact of ASR flexibility on RePtA VPP operation. (a) Adjustment period of 1 day. (b) Adjustment period of 3 days. (c) Adjustment period of 1 week. (d) Adjustment period of 2 weeks.}
        \label{Impact of ASR flexibility on RePtA VPP operation.}
    \end{figure}

    Furthermore, we consider the impact of reactor flexibility on energy costs. Fig. \ref{Impact of ASR flexibility on RePtA VPP operation.} shows the LCOA with the adjustment periods of the ASR from 1 day to 14 days.

    The benefits from the change in the hydrogen buffer are more significant than those of the ammonia buffer in Fig. \ref{Impact of ASR flexibility on RePtA VPP operation.}. In subfigure (a), the reduction from increasing the ammonia buffer is only \$2/t, while the reduction from increasing the hydrogen buffer is \$17/t. 
    
    Furthermore, the mean values of LCOA in the four subplots are \$220.79/t, \$229.15/t, \$234.96/t, and \$237.50/t, respectively. Reducing the adjustment time for the ASR from 14 days to 1 day results in a gain of approximately \$16.7/t, representing an approximately 7.0\% reduction in LCOA. The RePtA VPP operation becomes more flexible as the adjustment duration shortens, enabling more flexible bidding in the spot market.
 
    In Section \ref{Case Studies}, we introduce case studies based on an actual project in Inner Mongolia Autonomous Region. The results demonstrate the necessity of the hydrogen and ammonia buffer systems. Increasing the flexibility of the equipment and upgrading the storage capacity can help to reduce energy costs. Furthermore, hydrogen storage can be employed to balance the distribution within several hours, whereas the ammonia buffer is more suitable for addressing longer timescale price fluctuations. 

    \section{Conclusion}

    This paper proposes a multi-timescale trading strategy for an RePtA VPP in the electricity, hydrogen, and ammonia markets. Unlike previous studies, energy demand flexibility is provided by trading in the ammonia and hydrogen market. A two-stage robust optimization model is utilized to determine the multimarket trading, considering the uncertainty of renewable energy. The receding horizon optimization approach articulates the multiple timescales. Furthermore, case studies are carried out based on an actual project in Inner Mongolia Autonomous Region. The conclusions are as follows:

    1) By utilizing the hydrogen and ammonia buffer systems, the RePtA VPP can optimally coordinate production planning across hours and weeks to reduce energy costs. The hydrogen buffer is more effective for hourly price fluctuations, while the ammonia buffer is practical for longer-term price fluctuations.
    
    2) Increasing reactor flexibility enables a faster response of the RePtA VPP to market price fluctuations. Reducing the ASR adjustment period from 14 days to 1 day could result in a 7.0\% reduction in LCOA.

    Furthermore, with the widespread interest in the carbon market, its role in green hydrogen and ammonia trading has become increasingly prominent. The carbon market trading strategy of the RePtA VPP may be a promising direction for future research.

\end{document}